\newif\ifacm
\newif\ifusenix
\newif\ifneurips

\ifdefined\conftemplate\else
  \def\conftemplate{acm}
\fi
\expandafter\def\csname confswitch@acm\endcsname{\acmtrue}
\expandafter\def\csname confswitch@usenix\endcsname{\usenixtrue}
\expandafter\def\csname confswitch@neurips\endcsname{\neuripstrue}
\expandafter\ifx\csname confswitch@\conftemplate\endcsname\relax
  \PackageError{confswitch}{Unknown conference template `\conftemplate'}%
    {Set \string\conftemplate\space to acm, usenix, or neurips.}
\else
  \csname confswitch@\conftemplate\endcsname
\fi

\makeatletter
\def\input@path{{templates/acmart/}{templates/usenix/}{templates/neurips/}}
\makeatother

\newif\ifshellescape
\shellescapefalse

\ifneurips
  \documentclass{article}
  \PassOptionsToPackage{numbers,compress}{natbib}
  \usepackage[eandd]{neurips_2026}
\else\ifusenix
  \documentclass[letterpaper,twocolumn,10pt]{article}
  \usepackage{usenix}
\else
  \documentclass[sigplan,twocolumn]{acmart}
\fi\fi

\usepackage{confswitch}

\renewcommand\footnotetextcopyrightpermission[1]{}
\acmSubmissionID{\#28}
\usepackage{tikz}
\usepackage{amsmath}

\ifneurips
  \usepackage{amssymb,amsmath,amsfonts,amsthm}
  \usepackage[T1]{fontenc}
  \usepackage[utf8]{inputenc}
  \usepackage{microtype}
  \UseMicrotypeSet[protrusion]{basicmath}
  \usepackage{hyperref}
  \usepackage{graphicx}
  \usepackage[x11names]{xcolor}
  \usepackage{booktabs}
  \usepackage{etoolbox}
  \usepackage{comment}
  \usepackage{caption}
\else\ifusenix
  \usepackage{amssymb,amsmath,amsfonts,amsthm}
  \usepackage[T1]{fontenc}
  \usepackage[utf8]{inputenc}
  \usepackage{microtype}
  \UseMicrotypeSet[protrusion]{basicmath}
  \usepackage{hyperref}
  \usepackage{graphicx}
  \usepackage[x11names]{xcolor}
  \usepackage{booktabs}
  \usepackage{etoolbox}
  \usepackage{comment}
  \usepackage{caption}
\else
\fi\fi

\ifshellescape
  \usepackage{minted}
\else
  \usepackage{listings}   %% Tectonic-safe fallback while minted is off
\fi
\usepackage{upquote}
\usepackage{grffile}
\usepackage[normalem]{ulem}
\usepackage{multirow}
\usepackage{xspace}
\usepackage{tabularx}
\usepackage{ragged2e}
\usepackage{paralist}

\ifneurips
\else\ifusenix
  \usepackage[american]{babel}
\fi\fi

\ifneurips
  \usepackage{color}
\else\ifusenix
  \usepackage{color}   %% acmart loads xcolor; loading color after xcolor can conflict
\fi\fi
\usepackage{wrapfig}
\usepackage{balance}
\usepackage{enumitem}
\usepackage{epstopdf}
\usepackage{url}
\usepackage{pifont}
\usepackage{tikz}
\usepackage{mathtools}

\ifneurips
\else\ifusenix
  \usepackage{xkeyval}
\fi\fi

\usepackage{threeparttable}
\usepackage{makecell}

\ifneurips
  \usepackage[table,xcdraw]{xcolor}
\else\ifusenix
  \usepackage[table,xcdraw]{xcolor}
\fi\fi

\usepackage[tworuled, vlined]{algorithm2e}
\usepackage{sepfootnotes}
\ifneurips
\else\ifusenix
  \usepackage[compact]{titlesec}  %% acmart forbids \\section redefinition
\fi\fi
\usepackage[capitalize]{cleveref}
\crefformat{section}{§#2#1#3}
\usepackage{textcomp}
\usepackage{fancyvrb}
\usepackage[most]{tcolorbox}
\tcbuselibrary{listings,breakable}
\usepackage{subcaption}

\PassOptionsToPackage{usenames,dvipsnames}{color}
\hypersetup{unicode=true,
  colorlinks=true,
  linkcolor=blue,
  citecolor=blue,
  anchorcolor=blue,
  urlcolor=blue,
  breaklinks=true}
\setlist{topsep=1pt, partopsep=0pt, itemsep=1pt, parsep=1pt, leftmargin=10pt}

\makeatletter
\def\maxwidth{\ifdim\Gin@nat@width>\linewidth\linewidth\else\Gin@nat@width\fi}
\def\maxheight{\ifdim\Gin@nat@height>\textheight\textheight\else\Gin@nat@height\fi}
\makeatother
\setkeys{Gin}{width=\maxwidth,height=\maxheight,keepaspectratio}
\makeatletter
\g@addto@macro{\UrlBreaks}{\UrlOrds}
\makeatother

\newcommand\paraspace{\vspace*{0.4ex}}
\providecommand\parab[1]{\paraspace\noindent\textbf{#1}}

\newtoggle{reviewmode}
\newtoggle{anony}

\newcommand{\projurl}[1]{%
  \iftoggle{anony}{URL is hidden for review purpose}{\url{#1}}%
}

\newcommand{\ie}{\emph{i.e.,}\xspace}
\newcommand{\eg}{\emph{e.g.,}\xspace}

\newcommand{\secref}[1]{\S\ref{#1}}
\newcommand{\figref}[1]{Figure~\ref{#1}}
\newcommand{\tabref}[1]{Table~\ref{#1}}

\definecolor{teal}{rgb}{0.0, 0.5, 0.5}
\definecolor{olive}{rgb}{0.5, 0.5, 0.0}
\definecolor{pink}{rgb}{1.0, 0.75, 0.8}
\definecolor{lightgray}{gray}{0.75}
\definecolor{mediumgray}{gray}{0.5}
\definecolor{darkgray}{gray}{0.25}
\definecolor{charcoal}{gray}{0.2}
\definecolor{turquoise}{rgb}{0.25, 0.88, 0.82}
\definecolor{coral}{rgb}{1.0, 0.5, 0.31}
\definecolor{navyblue}{rgb}{0.0, 0.0, 0.5}
\definecolor{lime}{rgb}{0.75, 1.0, 0.0}
\definecolor{darkgreen}{rgb}{0.0, 0.5, 0.0}
\definecolor{violet}{rgb}{0.56, 0.0, 1.0}
\definecolor{lightgreen}{rgb}{0.85, 1.0, 0.85}
\definecolor{burgundy}{cmyk}{0.5, 1.0, 0.7, 0.4}
\definecolor{olivegreen}{cmyk}{0.64, 0, 0.95, 0.4}
\definecolor{peach}{cmyk}{0, 0.5, 0.7, 0}
\definecolor{mustard}{cmyk}{0, 0.3, 1, 0}



\newcolumntype{N}{>{\raggedleft\arraybackslash}p{1.4cm}}
\newcolumntype{U}{>{\raggedright\arraybackslash}p{0.8cm}}

\ifneurips
    \usepackage{titlesec}
    \titlespacing*{\section}{0pt}{7pt plus 3pt minus 3pt}{3pt plus 3pt minus 2pt}
    \titlespacing*{\subsection}{0pt}{4pt plus 3pt minus 2pt}{2pt plus 3pt minus 1pt}
    \titlespacing*{\subsubsection}{0pt}{4pt plus 3pt minus 2pt}{1pt plus 3pt minus 1pt}
    \titleformat{\section}{\large\bfseries}{\thesection}{1em}{}
    \titleformat{\subsection}{\normalsize\bfseries}{\thesubsection}{1em}{}
\else\ifusenix
    \usepackage{titlesec}
    \titlespacing*{\section}{0pt}{7pt plus 3pt minus 3pt}{3pt plus 3pt minus 2pt}
    \titlespacing*{\subsection}{0pt}{4pt plus 3pt minus 2pt}{1pt plus 3pt minus 1pt}
    \titlespacing*{\subsubsection}{0pt}{4pt plus 3pt minus 2pt}{0pt plus 3pt minus 1pt}
    \titleformat{\section}{\large\bfseries}{\thesection}{1em}{}
    \titleformat{\subsection}{\normalsize\bfseries}{\thesubsection}{1em}{}
  \fi\fi

\toggletrue{anony}  % disable author block
\newtoggle{showmarks}
\toggletrue{showmarks} % show comments and remarks
\iftoggle{showmarks}{
  \newcommand\red[1]{\textcolor{red}{#1}}
  \newcommand\redstrike[1]{\red{\sout{#1}}}
  \newcommand\green[1]{\textcolor{\green}{#1}}
  \newcommand\greenstrike[1]{\green{\sout{#1}}}
  \newcommand\orange[1]{\textcolor{orange}{#1}}
  \newcommand\orangestrike[1]{\orange{\sout{#1}}}
  \newcommand\blue[1]{\textcolor{blue}{#1}}
  \newcommand\bluestrike[1]{\blue{\sout{#1}}}
  \newcommand\purple[1]{\textcolor{purple}{#1}}
  \newcommand\purplestrike[1]{\purple{\sout{#1}}}
  \newcommand\teal[1]{\textcolor{teal}{#1}}
  \newcommand\tealstrike[1]{\teal{\sout{#1}}}
  \newcommand\turquoise[1]{\textcolor{turquoise}{#1}}
  \newcommand\turquoisestrike[1]{\turquoise{\sout{#1}}}
  \newcommand\darkgreen[1]{\textcolor{darkgreen}{#1}}
  \newcommand\darkgreenstrike[1]{\darkgreen{\sout{#1}}}
  \newcommand\lime[1]{\textcolor{lime}{#1}}
  \newcommand\limestrike[1]{\lime{\sout{#1}}}
  \newcommand\olivegreen[1]{\textcolor{olivegreen}{#1}}
  \newcommand\olivegreenstrike[1]{\olivegreen{\sout{#1}}}

  \newcommand{\yibo}[1]{[\purple{\sf\textit{#1 - Yibo}}]}
  \newcommand{\draft}[1]{\textcolor{turquoise}{\sf\textit{#1}}}
  \newcommand{\todo}[1]{[\textcolor{turquoise}{\sf\textbf{TODO: }\textit{#1}}]}
}{
  \newcommand\red[1]{#1}
  \newcommand\redstrike[1]{\unskip}
  \newcommand\green[1]{#1}
  \newcommand\greenstrike[1]{\unskip}
  \newcommand\orange[1]{#1}
  \newcommand\orangestrike[1]{\unskip}
  \newcommand\blue[1]{#1}
  \newcommand\bluestrike[1]{\unskip}
  \newcommand\purple[1]{\unskip}
  \newcommand\purplestrike[1]{\unskip}
  \newcommand\teal[1]{\unskip}
  \newcommand\tealstrike[1]{\unskip}
  \newcommand\turquoise[1]{\unskip}
  \newcommand\turquoisestrike[1]{\unskip}
  \newcommand\darkgreen[1]{\unskip}
  \newcommand\darkgreenstrike[1]{\unskip}
  \newcommand\lime[1]{\unskip}
  \newcommand\limestrike[1]{\unskip}
  \newcommand\olivegreen[1]{\unskip}
  \newcommand\olivegreenstrike[1]{\unskip}

  \newcommand{\yibo}[1]{}
  \newcommand{\draft}[1]{}
  \newcommand{\todo}[1]{}
}

\newcommand{\sysname}{\textsc{Hyperflux}\xspace}

\newcommand{\cmark}{\ding{51}}
\newcommand{\xmark}{\ding{55}}

\begin{document}

% \title{HyperFlux: Ultralight VM Substrate with Microsecond-Scale Cross-VM Core Elasticity}
\title{Offering Microsecond-Scale Cross-VM Core Elasticity on Colocated Lightweight Virtual Machines}

%% ============================================================
%% STEP 6: Authors (template-specific formatting)
%% ============================================================
\ifneurips
  \author{
    Author Name \\
    Department of Computer Science \\
    University or Institution \\
    \texttt{author@example.com}
  }
\else\ifusenix
    \author{
      {\rm Author Name}\\
      University or Institution
    }
  \else
    \author{Yibo Yan}
    \affiliation{
      \institution{University of Southern California}
      \city{Los Angeles}
      \country{USA}
    }
    % \email{author@example.com}
    \author{Seo Jin Park}
    \affiliation{
      \institution{University of Southern California}
      \city{Los Angeles}
      \country{USA}
    }
  \fi\fi

%% ============================================================
%% STEP 7: Abstract
%%   Keep the file name explicit so arXiv's static source scanner can find it.
%%   ACM collects the abstract before \maketitle; the other formats render it
%%   after \maketitle.
%% ============================================================
\ifacm
  \begin{abstract}
Serverless platforms commonly colocate many diverse workloads, each in a fast-booting, memory-lean virtual machine (VM), to improve deployment density.
Overprovisioning each VM for its peak protects tail latency during traffic bursts but hurts density; 
maintaining high density while effectively protecting tail latency requires the infrastructure to be able to shift physical cores, at a microsecond timescale, to whichever latency-sensitive VM is bursting and reclaim them as the burst subsides.
No VM substrate delivers this: conventional VMs resize a guest's cores only through a millisecond-scale vCPU hot-plug path, Firecracker fixes a VM's core count at boot, and the ultralight VMs that boot fastest drop multicore execution entirely.

We present \sysname, a commodity-KVM ultralight VM substrate that makes a VM's \emph{parallelism width} (the number of physical cores backing it) elastic at runtime.
We show that \sysname can move a core across VMs in merely 13\,$\mu$s, even when forcibly reclaiming it from a busy donor, orders of magnitude faster than vCPU hot-plug.
A \sysname VM incurs only a 3.2\,MB memory footprint and can cold-boot in 1.37\,ms, on par with the fastest-booting ultralight VMs, while uniquely supporting multicore parallelism.
Under colocation, it can reduce high-priority VMs' tail latency by up to 10x under high load compared to static core-sharing with Firecracker and Cloud Hypervisor, and deliver a lower and more stable tail latency compared to using cgroup and vCPU hot-plug under changing load bursts.
\end{abstract}

\fi

%% ============================================================
%% STEP 8: CCS concepts (ACM-only; ignored by USENIX/NeurIPS)
%% ============================================================

% \keywords{}

%% ============================================================
%% STEP 9: \confmaketitle handles ordering correctly
%% ============================================================
\ifusenix\date{}\fi
\confmaketitle
\ifacm\else
  \begin{abstract}
Serverless platforms commonly colocate many diverse workloads, each in a fast-booting, memory-lean virtual machine (VM), to improve deployment density.
Overprovisioning each VM for its peak protects tail latency during traffic bursts but hurts density; 
maintaining high density while effectively protecting tail latency requires the infrastructure to be able to shift physical cores, at a microsecond timescale, to whichever latency-sensitive VM is bursting and reclaim them as the burst subsides.
No VM substrate delivers this: conventional VMs resize a guest's cores only through a millisecond-scale vCPU hot-plug path, Firecracker fixes a VM's core count at boot, and the ultralight VMs that boot fastest drop multicore execution entirely.

We present \sysname, a commodity-KVM ultralight VM substrate that makes a VM's \emph{parallelism width} (the number of physical cores backing it) elastic at runtime.
We show that \sysname can move a core across VMs in merely 13\,$\mu$s, even when forcibly reclaiming it from a busy donor, orders of magnitude faster than vCPU hot-plug.
A \sysname VM incurs only a 3.2\,MB memory footprint and can cold-boot in 1.37\,ms, on par with the fastest-booting ultralight VMs, while uniquely supporting multicore parallelism.
Under colocation, it can reduce high-priority VMs' tail latency by up to 10x under high load compared to static core-sharing with Firecracker and Cloud Hypervisor, and deliver a lower and more stable tail latency compared to using cgroup and vCPU hot-plug under changing load bursts.
\end{abstract}

\fi
\ifneurips\else
  \pagestyle{plain}
\fi

%% ============================================================
%% Body sections
%%   Add your \input{sections/...} lines here.
%% ============================================================

\section{Introduction}
\label{sec:intro}

Serverless platforms now routinely run computations in microVMs, which pair the hardware-enforced isolation of a virtual machine (VM) with the fast cold startup and small memory footprint~\cite{agachefirecrackerlightweightvirtualization2020}.
This reflects a decade-long progression in virtualization in the context of the serverless computing landscape: from conventional VMs (\eg QEMU~\cite{bellard2005qemu}), to lightweight VMs (\eg Cloud Hypervisor~\cite{2026cloud}, Kata~\cite{2026kata}), to microVMs (\eg Firecracker~\cite{agachefirecrackerlightweightvirtualization2020}) and unikernels~\cite{madhavapeddyunikernelsrisevirtual2013, kuenzer2021unikraft}, and most recently to \emph{ultralight} VMs such as Dandelion~\cite{kuchler2025unlocking} and Hyperlight~\cite{penna2026hyperlight} that can cold-start in milliseconds on commodity hardware.
Each step traded functionality for agility, shedding devices, kernel surface, and OS services to boot faster and minimize memory footprint.
The payoff is density: providers can theoretically pack more such instances onto one host~\cite{li2022rund}, amortizing the hardware across more tenants.

However, raw density does not translate into high utilization when bursty, latency-sensitive serverless workloads exist~\cite{shahradserverlesswildcharacterizing2020}.
These workloads run across multiple cores with time-varying parallelism, such as request-serving microservices, web servers, media transcoding, and realtime data-parallel analytics; their core demand swings sharply under tight Service-Level Objectives (SLOs), suddenly needing many cores to absorb a burst of core demands, then falling idle between bursts.
% A single-threaded function gains nothing from an extra core and is outside our scope.
This class is now mainstream in commercial serverless analytics and AI, spanning serverless data analytics and warehouses~\cite{2026compute, 2026sql, 2025architecting, 2026amazon, 2026snowflake, autoscale}, serverless ML inference and compute~\cite{deploy, 2026modal, 2026configure}, and multi-core serverless containers~\cite{2026instance, 2026serverless}; 
cloud providers autoscale each workload instance's parallelism under load and downscale it to zero-core when idle, so an instance both re-starts and re-scales frequently, exercising fast startup and elastic parallelism over its lifetime.
To protect the tail latency of these high-priority bursty services from core contention with colocated low-priority VMs, the provider often resorts to overprovisioning their VMs for peak demand, thereby stranding the very capacity that density was meant to reclaim~\cite{dean2013tail, demoulin2021when, hall2026harvesting}.

\begin{figure}
    \centering
    \includegraphics[width=\linewidth]{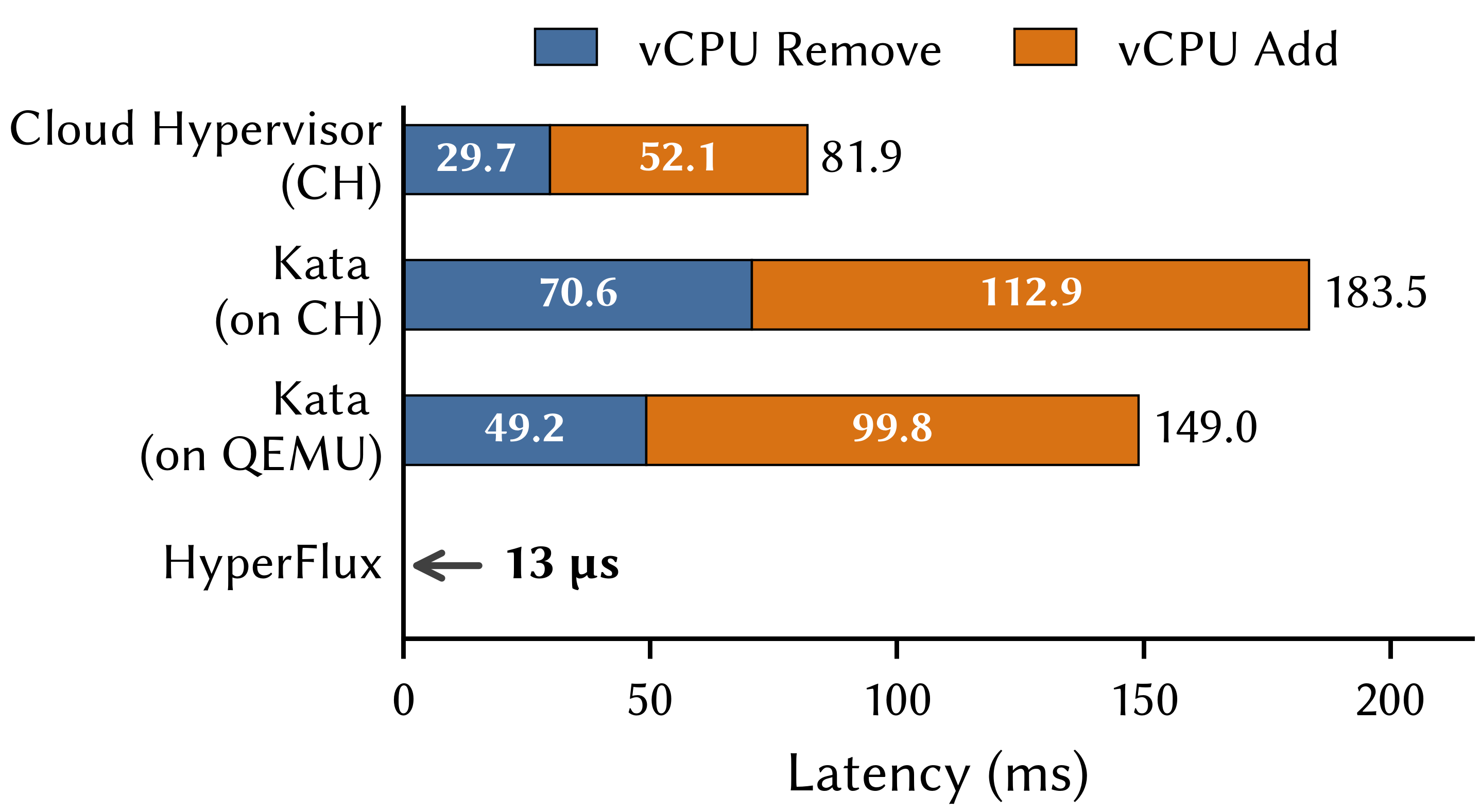}
    \caption{Moving a core between colocated VMs with conventional vCPU hot-plug (remove from the donor and add to the recipient). The full core movement (remove and add) in the conventional path takes hundreds of milliseconds, while \sysname takes only 13\,$\mu$s.
    }
    \label{fig:core-realloc}
\end{figure}

Many systems use \textit{core harvest} either at coarse timescales or designed for conventional heavyweight VM substrates, shifting idle cores to high-priority VMs and reclaiming them as demand falls~\cite{stojkovic2025hardharvest, ambati2020providing, wang2021smartharvest}.
No VM substrate shifts cores promptly enough at the finer timescale bursts demand, protecting these services' tail latency requires reacting in \textit{tens of microseconds}~\cite{hall2026harvesting}, while preserving ultralight VM properties that modern serverless computing requires.
Lightweight VMs~\cite{2026cloud,2026kata} resize cores through the ACPI CPU hot-add/remove path~\cite{cpuhotplugkernel}, but as \figref{fig:core-realloc} shows, a single hot-add or hot-remove takes milliseconds, orders of magnitude beyond that timescale.
Firecracker fixes a VM's vCPU count at boot, offering no core harvest at all, and the fastest-booting ultralight VMs (\eg Dandelion~\cite{kuchler2025unlocking}, Hyperlight~\cite{penna2026hyperlight}) strip out threading and multicore entirely, so a donated core would have no parallel work to run.
\looseness=-1

We argue that one can design an ultralight VM substrate that delivers all three desired properties simultaneously:
\begin{inparaenum}[(i)]
    \item millisecond-scale cold-start latency on commodity hardware with small memory footprint;
    \item support for threading and multicore so that VMs can harness local cores immediately for parallelism on-demand;
    \item integrated core harvest support in the substrate (in Virtual Machine Monitor, or VMM) that can shift cores at microsecond scale among colocated VMs to protect high-priority latency-sensitive workloads during bursts without squandering core resources on overprovisioning for the peak demand.
\end{inparaenum}

We present \sysname, a commodity KVM-based ultralight VM substrate that makes a VM's \emph{parallelism width} (the number of physical cores backing it) elastic at runtime, enabling microsecond-scale cross-VM core harvest and redistribution.
To achieve this without sacrificing agility, \sysname tightly co-designs three VM substrate components from the ground up:
\begin{inparaenum}[(i)]
    \item \textsc{Fluxion}, a virtual machine monitor (VMM) that manages VMs and also serves as a host-side core arbiter that decides when and where to move cores from live demand signals;
    \item \textsc{kFlux}, a kernel module that actuates core movements in microseconds; and
    \item \textsc{FluxOS}, a guest library OS (LibOS) that provides multicore execution with threading and,
    % while perserving ultralight VMs properties with naive support for microsecond-scale core movements.
    on a reclaim request, preemptively parks the running guest thread without relying on cooperative scheduling so that the host can take cores promptly.
\end{inparaenum}
\sysname keeps the ultralight VM virtues, preserving a fast startup speed, a small memory footprint, and hardware-enforced isolation, while enabling cores to \emph{flux} between colocated VMs as demand moves at $\mu$s-scale.

In our evaluation, \sysname moves a core between VMs in about 13\,$\mu$s even when forcibly reclaiming it from a busy donor (7.4\,$\mu$s to free the donor's core and 5.6\,$\mu$s to install it on the recipient), orders of magnitude faster than ACPI vCPU hot-plug (\textasciitilde100--200\,ms).
Under colocation, \sysname promptly shifts cores to whichever high-priority VM is bursting and reclaims them as load subsides, holding high-priority p99 latency more than an order of magnitude below static core-sharing with Firecracker and Cloud Hypervisor.
Under continuously changing burst loads, \sysname maintains a lower and more stable tail latency for high-priority VMs than oracle-driven \texttt{cgroup} and vCPU hot-plug under bursts, while staying on par with modern ultralight VMs on memory footprint and cold-start latency.

This paper makes the following contributions:
\begin{itemize}
    \item We identify \emph{elastic parallelism width}, which requires host-driven, cross-VM control of the cores backing a VM at runtime, as a missing axis in the VM design space (\secref{sec:missingaxis}).
    \item We present the design of \sysname, the first commodity-KVM ultralight VM substrate to realize microsecond-level elastic parallelism width and cross-VM core movements, and design rationales behind three co-designed components: \textsc{Fluxion} (VMM), \textsc{kFlux} (a kernel module), and \textsc{FluxOS} (guest OS) (\secref{sec:design}).
    \item We present an evaluation that showcases cross-VM core handoff can be achieved in $\sim$13\,$\mu$s and such a mechanism can effectively protect tail latency compared to both static core-sharing and dynamic adjustment with cgroup and vCPU hot-plug.
    Meanwhile, \sysname retains the fast startup speed and low memory footprint that are on par with most modern ultralight VMs (\secref{sec:eval}).
\end{itemize}

\section{Background}

\subsection{Library OS (LibOS)}
A \emph{library OS} (LibOS) is an OS abstraction layer delivered as a library that an application links against, running the application and its OS services in a single address space with no kernel/user boundary~\cite{englerexokerneloperatingsystem1995, porterrethinkinglibraryos2011}.
Its service surface is a choice rather than fixed minimalism, from a near-complete POSIX API down to a bare runtime~\cite{kuenzer2021unikraft}.
Most modern ultralight VMs take the LibOS approach but slim that surface aggressively to boot faster with minimal memory footprint~\cite{madhavapeddyunikernelsrisevirtual2013, kuchler2025unlocking, penna2026hyperlight}.
\sysname is co-designed with a guest LibOS, \textsc{FluxOS}, that retains these ultralight advantages while adding a lightweight userspace thread abstraction and scheduler that can preempt a running thread and relinquish a core to the host in microseconds.

\subsection{Virtual CPU (vCPU) Abstraction}
% \yibo{Explain the core concept of vCPU is running as a thread and can be scheduled to cores.}
Under KVM~\cite{kivity2007kvm}, each virtual CPU (vCPU) a guest VM sees is backed by an ordinary host kernel thread that the host kernel schedules onto a physical core; 
the guest vCPU advances only while that thread is running on a core. 
A guest is typically assigned a fixed number of vCPUs at boot, but this number is merely its \emph{logical} view of how many processors it has.
How many of those vCPU hosting kernel threads occupy physical cores at any instant is a separate decision made entirely on the host side.
A guest's active parallelism is thus bounded by its vCPU count but determined, at different moments, by how many of its vCPU threads the host places on cores. % VERIFY: HyperFlux varies the *physical backing* (parks/unparks vCPU threads) while keeping the guest's vCPU count fixed -- confirm before merging.

\parab{The host controls vCPUs at \texttt{VMExit}s.}
Each vCPU thread drives the guest through a tight loop: an \texttt{ioctl(KVM\_RUN)} enters guest mode and executes guest instructions directly on the core until a \texttt{VMExit}, which can be triggered by I/O, an interrupt, a hypercall, or a signal delivered to the hosting kernel thread.
Upon a \texttt{VMExit}, KVM returns vCPU control to the host, which services the event and re-enters.
As the guest owns the core between \texttt{VMExit}s, this control loop is the host's only lever on a running vCPU. 
Specifically, to act on one that is busy executing guest code, the host must first force it to exit.

% \parab{Reclaiming a core na\"ively can strand the guest's work.}
% Taking a core back, in principle, is essentially descheduling its vCPU thread from the physical cores, but how the host does so decides whether the guest's work will be stranded or not.
% If the host simply lets its scheduler preempt the thread, the core frees at once, yet any guest thread caught mid-execution is frozen inside that vCPU's saved register and stack state---no other vCPU can resume it, so the work is \emph{stranded} until the same vCPU runs again.
% Avoiding this requires coordinating with the guest beforehand, so that in-flight work is handed to a surviving vCPU before the core is taken.
% The naive path is fast but unsafe under load; the safe path, when routed through the guest kernel's hot-plug machinery, is slow.
% \sysname's design (\S\ref{sec:design}) turns on closing exactly this gap---preempting and rescuing the in-flight work in microseconds---whereas conventional hot-plug pays a millisecond cost to reconfigure the guest's own view of its CPU set (\S\ref{subsec:designspace}).

% \section{Limitations of Existing MicroVM Substrates and Mechanisms}
\section{The Missing Axis: Microsecond Cross-VM Core Elasticity}
\label{sec:missingaxis}

% \sysname targets three properties at once: an ultralight VM infrastructure, true multicore execution, and microsecond cross-VM core elasticity.
% No existing substrate delivers all three: colocation demands controlled core movement (\S\ref{subsec:contention}), yet every available mechanism fails on at least one axis (\S\ref{subsec:gauntlet}), leaving a gap in the commodity-KVM design space that \sysname is the first to fill (\S\ref{subsec:designspace}).

% \subsection{Static Allocation with Uncontrolled Core Sharing Creates Contentions}
\subsection{Static Allocation with Uncontrolled Core Sharing Creates Contention}
\label{subsec:contention}

The simplest way to raise utilization under colocation is to make several VMs share one set of physical cores under the host (Linux) scheduler, but this fails for latency-sensitive workloads: under bursts the scheduler is oblivious to demand and, aiming for fair sharing, has no mechanism to direct cores, inflating tail latency for high-priority services~\cite{zhangcpi2cpuperformance2013, iorgulescu2018perfiso, dean2013tail}.
Statically partitioning cores for peak demand instead avoids core contention under bursts but strands resources during idle periods~\cite{demoulin2021when, hall2026harvesting}.
Escaping this dilemma requires neither static partitioning nor uncontrolled sharing but \emph{controlled} core movement: shifting a core to whichever guest needs it and reclaiming it when the burst passes, at the timescale bursts demand.

\subsection{No Existing Mechanism Moves Cores Promptly Across VMs}
\label{subsec:gauntlet}

We walk through the mechanisms an operator could reach for. Each fails on at least one of \sysname's three axes.

\parab{vCPU hot-plug is millisecond-scale.}
Cloud Hypervisor~\cite{2026cloud} can change a guest's vCPU count, but only by driving the guest kernel's ACPI CPU offline/online path~\cite{cpuhotplugkernel}: the guest VM must quiesce the target CPU, migrate its tasks away, and tear down per-CPU state before the host can repurpose the core.
As we discussed earlier in \secref{sec:intro}, this process takes hundreds of milliseconds to finish, rendering such core harvest ineffective for bursty services.

\parab{Linux \texttt{cgroup} arbitrates shares, not allocation.}
As a vCPU is hosted by a regular host thread, Linux may use the \texttt{cgroup} cpu subsystem~\cite{controlgroupv2} to control the resource usage.
The \texttt{cgroup} cpu subsystem can throttle a guest's vCPU threads, but it regulates each thread's \emph{time share} rather than granting a thread exclusive use of a reclaimed core.
A recent study of CPU harvesting in container systems also shares the same insight: share-based controls leave significant residual interference on latency-sensitive instances and also react slowly (order of milliseconds) to absorb the demand spike~\cite{hall2026harvesting}.

\parab{Production harvesting is coarse and asymmetric.}
Cluster harvesting does move cores across VMs: Azure's resource-harvesting VMs~\cite{ambati2020providing} and SmartHarvest~\cite{wang2021smartharvest} reclaim a primary VM's idle cores for a colocated batch VM.
But the movement is \emph{asymmetric} (it shields one latency-critical primary while feeding best-effort batch, rather than balancing peer guests), takes only \emph{already-idle} cores, and reacts far more slowly than a microsecond burst, on conventional heavyweight VMs.
It improves batch throughput, not the tail-bounded utilization of colocated latency-critical guests.

% \parab{Microsecond core schedulers stop at the VM boundary.}
% Within a single OS, schedulers such as Caladan and Shenango~\cite{fried2020caladan, ousterhout2019shenango} reallocate cores among cooperating tasks in microseconds---demonstrating that the timescale is achievable.
% But they operate inside one trust domain and cannot move a core across the VM isolation boundary, where donor and recipient share neither address space nor scheduler.
% Elastic-vCPU designs that do reach into a guest (\eg Ditto~\cite{zhao2024ditto}) remain bounded by cooperative guest check-in intervals, and hardware proposals (HardHarvest~\cite{stojkovic2025hardharvest}) require new silicon rather than commodity KVM.

\parab{Ultralight VMs do not have in-VM multi-core parallelism.}
Finally, the substrates that best satisfy the lightweight axis defeat the multicore one: the VMs that boot fastest and pack densest, such as Dandelion~\cite{kuchler2025unlocking} and Hyperlight~\cite{penna2026hyperlight}, commonly strip out threading and multi-core support. % VERIFY: per-system multicore status (paper_anchor.md open item 2)
Efforts to restore it underscore the gap: Nanvix~\cite{segarra2026nanvix} adds threading atop Hyperlight, but multiplexes those threads onto a single vCPU, supporting concurrency without true parallelism.
For all of these, prompt core harvest is not meaningful: a guest that runs its work on a single vCPU gains nothing from a donated core.

% =========================================================================
% tables/position.tex -- design-space / positioning table for S2 (sec:missingaxis).
% Include with \input{tables/position} inside a \subsection, and reference as
% \tabref{tab:position}  (replaces the old \figref{fig:design-space}).
%
% Columns ARE the requirements derived in the motivation (so the matrix is not
% cherry-picked): Lightweight | True multicore | Cross-VM | us-scale.
% Scope: SOFTWARE substrates only (HardHarvest = new HW, omitted; Caladan/Shenango
% = not VMs, omitted -> both covered in S7 Related Work prose).
% "Cross-VM" and "us-scale" are split on purpose: the empty corner = only HyperFlux
% is checked on BOTH. Caveats (asymmetric / check-in / not-a-VM) -> footnotes.
%
% Needs (all already in 0_packages.tex): booktabs, pifont, threeparttable,
% makecell, colortbl. \cmark/\xmark from 0_macros.tex; lightgreen from 0_macros.tex.
% If it overflows \columnwidth, switch `table`->`table*` (full width) or wrap the
% tabular in \adjustbox{max width=\columnwidth}{...}.
% =========================================================================

\providecommand{\pmark}{\textcolor{mediumgray}{$\sim$}} % partial / qualified
\providecommand{\namark}{\textcolor{mediumgray}{--}}     % not applicable
% \rowcolor needs xcolor's [table] option (colortbl). 0_packages.tex loads it for
% NeurIPS/USENIX but NOT in ACM mode. Guard so the table compiles everywhere:
% where [table] is present the HyperFlux row is highlighted; in ACM it degrades to
% no fill (the row is still bold). To enable the fill in ACM too, add
% \PassOptionsToPackage{table}{xcolor} at the very top of 1_main.tex (before \documentclass).
\providecommand{\rowcolor}[1]{}

% Muted mark colors -- softer than stock red/green, defined once (idempotent).
\providecolor{markgreen}{HTML}{2E7D52} % muted forest green
\providecolor{markred}{HTML}{C0392B}   % muted brick red
\begin{table}[t]
  \centering
  % Colored marks, scoped to this table (the table environment forms a group, so
  % these \renewcommands do not leak to the rest of the document). \pmark stays gray.
  \renewcommand{\cmark}{\textcolor{markgreen}{\ding{51}}}
  \renewcommand{\xmark}{\textcolor{markred}{\ding{55}}}
  \footnotesize
  \setlength{\tabcolsep}{2pt}
  \renewcommand{\arraystretch}{1.15}
  \begin{threeparttable}
    {\footnotesize\cmark~Supported\hspace{1.5em}\pmark~Partial\hspace{1.5em}\xmark~Not Supported\par}
    \vspace{3pt}
    \begin{tabular*}{\columnwidth}{@{\extracolsep{\fill}}l cccc@{}}
      \toprule
      \textbf{Substrate} &
      \makecell{Light\\weight} &
      \makecell{True\\Multicore} &
      \makecell{Cross\\VM} &
      \makecell{$\mu$s\\scale} \\
      \midrule
      Firecracker~\cite{agachefirecrackerlightweightvirtualization2020}        & \cmark & \cmark & \xmark & \xmark \\
      Cloud Hypervisor\,/\,Kata~\cite{2026cloud,2026kata}                                & \pmark & \cmark & \xmark\tnote{$\dagger$} & \xmark\tnote{$\dagger$} \\
      Dandelion~\cite{kuchler2025unlocking}, Hyperlight~\cite{penna2026hyperlight} & \cmark & \xmark & \xmark & \xmark \\
      Nanvix~\cite{segarra2026nanvix}                                          & \cmark & \xmark & \xmark & \xmark \\
      Harvest VMs~\cite{ambati2020providing,wang2021smartharvest}              & \xmark & \cmark & \cmark & \xmark \\
    %   Caladan\,/\,Shenango~\cite{fried2020caladan,ousterhout2019shenango}      & \namark\tnote{e} & \cmark & \xmark\tnote{e} & \cmark \\
      Ditto~\cite{zhao2024ditto}                                               & \pmark\tnote{$*$} & \cmark & \pmark\tnote{$*$} & \xmark\tnote{$*$} \\
      \midrule
      \rowcolor{lightgreen}
      \textbf{\sysname}                                                        & \cmark & \cmark & \cmark & \cmark \\
      \bottomrule
    \end{tabular*}
    \begin{tablenotes}\footnotesize
      \item[$\dagger$] Per-VM ACPI hot-plug only, not a native cross-VM core handoff.
      \item[$*$] Confidential-VM runtime, heavier than a microVM; Cross-VM core movement requires cooperative scheduling and a specific paradigm.
    \end{tablenotes}
    \caption{Where existing substrates fall on the requirements for microsecond cross-VM core elasticity (\secref{sec:missingaxis}): a lightweight substrate, true multicore execution, and core movement that is both cross-VM and at {$\mu$s-scale}. 
    Each prior substrate misses at least one; only \sysname{} satisfies all.
    }
    \label{tab:position}
  \end{threeparttable}
\end{table}

\subsection{The Missing Axis and \sysname's Position}
\label{subsec:designspace}

These failures trace two coupled dimensions.
\emph{Elastic parallelism width} concerns a guest in isolation: it runs across multiple cores with true parallelism and its core demand can change sharply at runtime.
The existing ultralight and static substrates lack this property.
\emph{Cross-VM core elasticity} concerns colocated guests on a shared host: cores move promptly across the isolation boundary from one guest to another, which hot-plug and Linux built-in resource controllers cannot achieve effectively at microsecond scale.
The two are inseparable on a fixed-size host server: a guest's parallelism width can only grow by harvesting a neighbor's core, and a harvested core only helps a recipient that can absorb it.

\tabref{tab:position} makes the gap concrete across the requirements these two dimensions imply.
No prior substrate satisfies all of them: each is either too heavy, limited to a single core, unable to move cores across the VM boundary, or too slow to do so at the timescale bursts demand.
\sysname is the first commodity-KVM substrate to satisfy all four, and the only one that does so without trading away the lightweight, multicore execution the others sacrifice.
% NOTE: positions HyperFlux CONCEPTUALLY (the empty cell). Mechanism-by-mechanism
% comparison lives in S7 Related Work -- keep them non-overlapping.

% \yibo{Discuss and show static allocation with firecracker that shared the physical cores are not working.}

% \subsection{Existing Core Elasticity Are too Slow}

% \yibo{Show Cloud Hypervisor maybe}

% \subsection{Limitation with Linux Kernel-Level Resource Control}
% \yibo{Try out using cgroup to share the vCPU on physical cores.}
\section{\sysname: Enabling Cross-VM Core Elasticity at Microsecond-Scale}
\label{sec:design}

% \subsection{Design Objectives and Positioning}

\begin{figure}
    \centering
    \includegraphics[width=\textwidth]{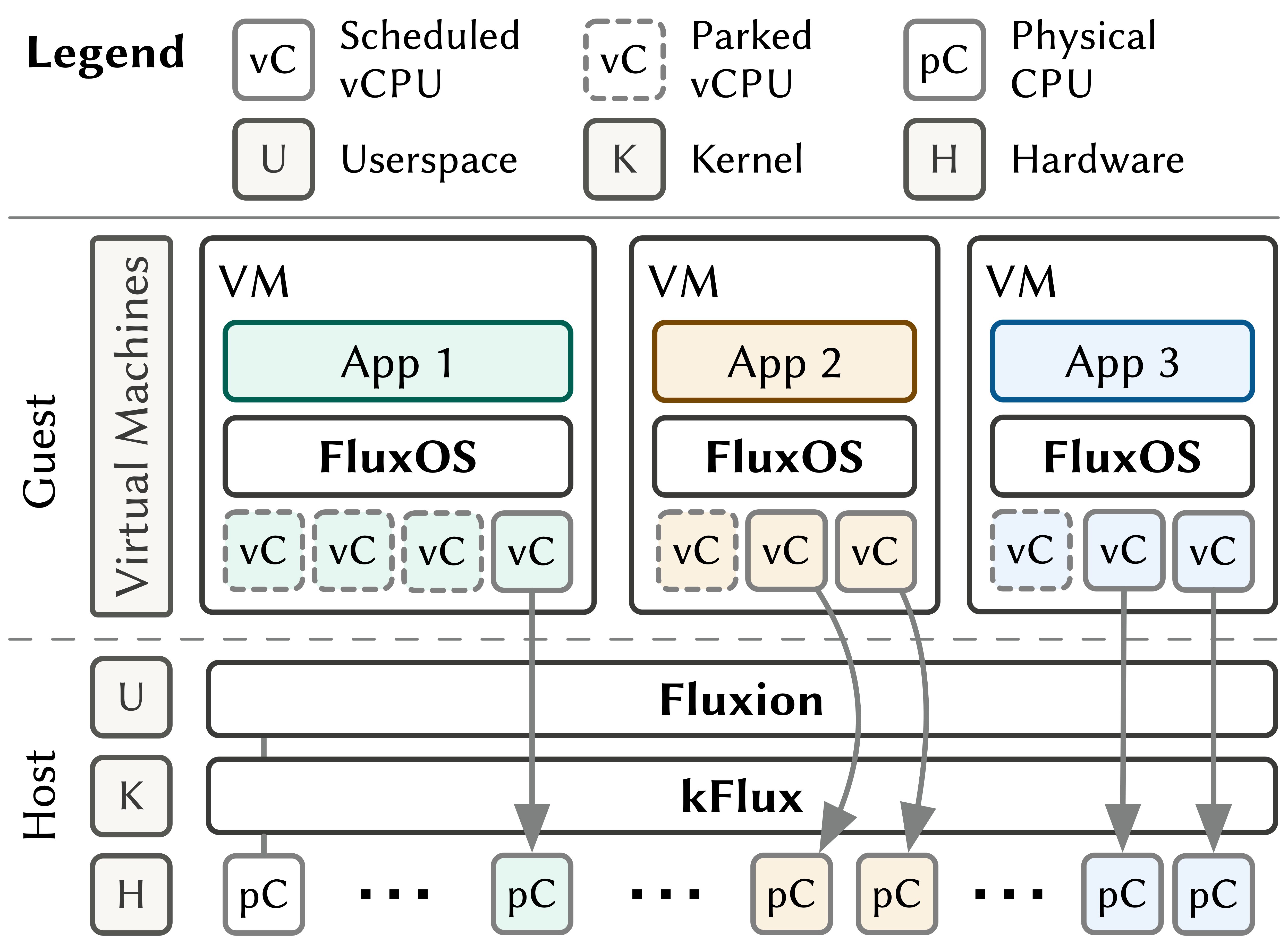}
    \caption{High-level architecture overview of \sysname.}
    \label{fig:arch}
\end{figure}

\parab{Core Enabler.}
\sysname makes a VM's \emph{parallelism width}, the number of physical cores backing it, a quantity the host or Virtual Machine Monitor (VMM) adjusts at runtime.
This rests on how VMs run on commodity KVM: a guest's vCPU count is only a logical view, since each vCPU is an ordinary host thread and the host alone decides how many of those threads occupy a physical core at any instant (\secref{sec:missingaxis}).
\sysname performs a fast runtime action guided by a core allocation policy: it parks the thread behind one vCPU and unparks another to grow, shrink, and shift a VM's width across colocated VMs at microsecond scale.
Therefore, \sysname does not change the guest's configured vCPU count at runtime, avoiding a guest-side ACPI CPU hot-plug-like control path.
As a result, the guest does not need to reconfigure its view of core topology; instead, it only needs to be aware that a vCPU is not progressing (parked).
The provider allows tenants to configure the maximum parallelism, in terms of core count, that instances can scale up to;
and the host VMM makes the harvest and scaling decisions according to the core demand and configured priority class.
Because \sysname adjusts the physical backing of a fixed set of vCPU threads, elastic parallelism width is a property of the host's vCPU management, not the guest ABI: any runtime that publishes demand signals and honors park/unpark can use it, and \textsc{FluxOS} is one such runtime.
% Holding the guest's CPU set fixed is what collapses a core move from the milliseconds an ACPI offline/online cycle costs (\figref{fig:core-realloc}) to microseconds, since neither guest reconfigures its view of the machine.

\parab{Challenges.}
The core challenge is that \sysname must perform core movements \emph{across the VM isolation boundary}, among mutually untrusted guests that share neither an address space nor a scheduler with the host or with one another.
Doing so raises three problems that core scheduling within a single trusted domain~\cite{ousterhout2019shenango,fried2020caladan} never confronts.
First, the host's only schedulable handle is the vCPU thread, not the guest threads and run-queues inside it.
The trust surface between host and guest should be narrow in that the host has no direct view into guest schedulers and run-queues, whereas a single-domain scheduler like Caladan~\cite{fried2020caladan} directly reads scheduler internals via memory mapping.
% \sysname must agree with each guest on a narrow shared-memory channel for core demand and control, and treat what it reads there as untrusted (\secref{subsec:controlplane}).
Second, the host can stop a vCPU but cannot rescue the work it was running, which lives in guest state only the guest's runtime can save and migrate.
Reclaiming a core therefore requires the guest runtime itself to preempt and save the in-flight work so that a surviving vCPU may pick it up to maximally avoid work stranding.
Third, we should preserve the ultralight property (\ie fast startup latency and minimal memory overhead), despite additional mechanisms for dynamic parallelism width and cross-VM core movement.
% Third, because the guest is untrusted and may be slow or adversarial, the host cannot assume a reclaim request is honored promptly; every core move must be a bounded, revocable transaction rather than a fire-and-forget signal (\secref{subsec:reclaim}).

% First, the control plane must cross the isolation boundary to reach an untrusted guest with which the host shares no memory and no scheduler (\secref{subsec:controlplane}).
% Second, reclaiming a core may require preempting a vCPU in mid-execution and rescuing its in-flight work without stranding it, because the workload may never reach a yield point (\secref{subsec:reclaim}).
% Third, because a guest may be slow or adversarial, every core move must be a bounded, revocable transaction rather than a fire-and-forget signal (\secref{subsec:reclaim}).

\parab{Architecture Overview.}
\figref{fig:arch} shows the resulting architecture, in which \sysname co-designs three provider-controlled components that form a closed loop.
\textsc{FluxOS}, a guest LibOS, runs the workload across lightweight guest threads (gThreads) and publishes each VM's live core demand; on a reclaim request it preemptively parks the running gThread so a sibling vCPU resumes it (\secref{subsec:reclaim}).
\textsc{Fluxion}, a host user-space VMM, decides when and where to move a core, and \textsc{kFlux}, a host kernel module, actuates the move by pinning and unpinning vCPU threads in microseconds without a system call or IPI on the fast path.

\parab{Confidential VM.}
The design extends to confidential VMs~\cite{2026amd, kuvaiskii2024graminetdx, 2026intel} (CVM) without redesign: it needs no host access to guest memory beyond the existing KVM interface and a narrow shared-memory region, which a bounce buffer can provide.
The full CVM design is a potential future work.
% \looseness=-1

\subsection{Threat Model}
\label{subsec:threat}

\sysname's host-side components, \textsc{Fluxion} and \textsc{kFlux}, are provider-controlled and trusted, exactly as the VMM and host kernel are in a conventional microVM deployment.
\textsc{FluxOS} is the provider-supplied guest substrate; it shares the guest's address space with the tenant workload and is not hardware-isolated from it, so none of \sysname's host-side guarantees depend on \textsc{FluxOS}'s integrity, only on the host's control of the vCPU thread and on hardware isolation.
The tenant workload linked against \textsc{FluxOS} is untrusted and may be buggy or behave adversarially.
Tenant isolation is hardware-enforced by KVM~\cite{kivity2007kvm} and is unchanged from a stock microVM: \sysname adds a per-VM shared-memory control plane between each guest and the host but never a channel between guests, so no tenant state crosses the isolation boundary unless guests intentionally communicate (\eg via the network).
% \looseness=-1

Two properties follow that the rest of the design depends on.
First, the host can always reclaim a core from a non-yielding workload: a vCPU is an ordinary host thread the host can force out of guest execution.
Second, a guest can influence only its own scheduling, never the host's liveness or its neighbors'.
A VM's priority class and guaranteed core floor are provider-set and live in host state the guest cannot reach.
The core allocator never blocks on a guest and never reclaims a neighbor below its guaranteed floor, so a slow or malicious guest can forfeit its own cores but cannot stall the allocator or push a neighbor below its guarantee.

A malicious VM can retain more cores than it actually needs by generating phantom demands.
Such a tenant gains nothing because the core usage is metered and billed, and a VM can never exceed its provider-configured maximum parallelism width.
Persistent core capacity contention created by such events should be handled one layer up, where the provider's placement layer can relocate a VM that will not release cores onto an uncontended host.

\subsection{FluxOS: an Ultralight Multicore LibOS}
\label{subsec:fluxos}

\begin{figure}
    \centering
    \includegraphics[width=\columnwidth]{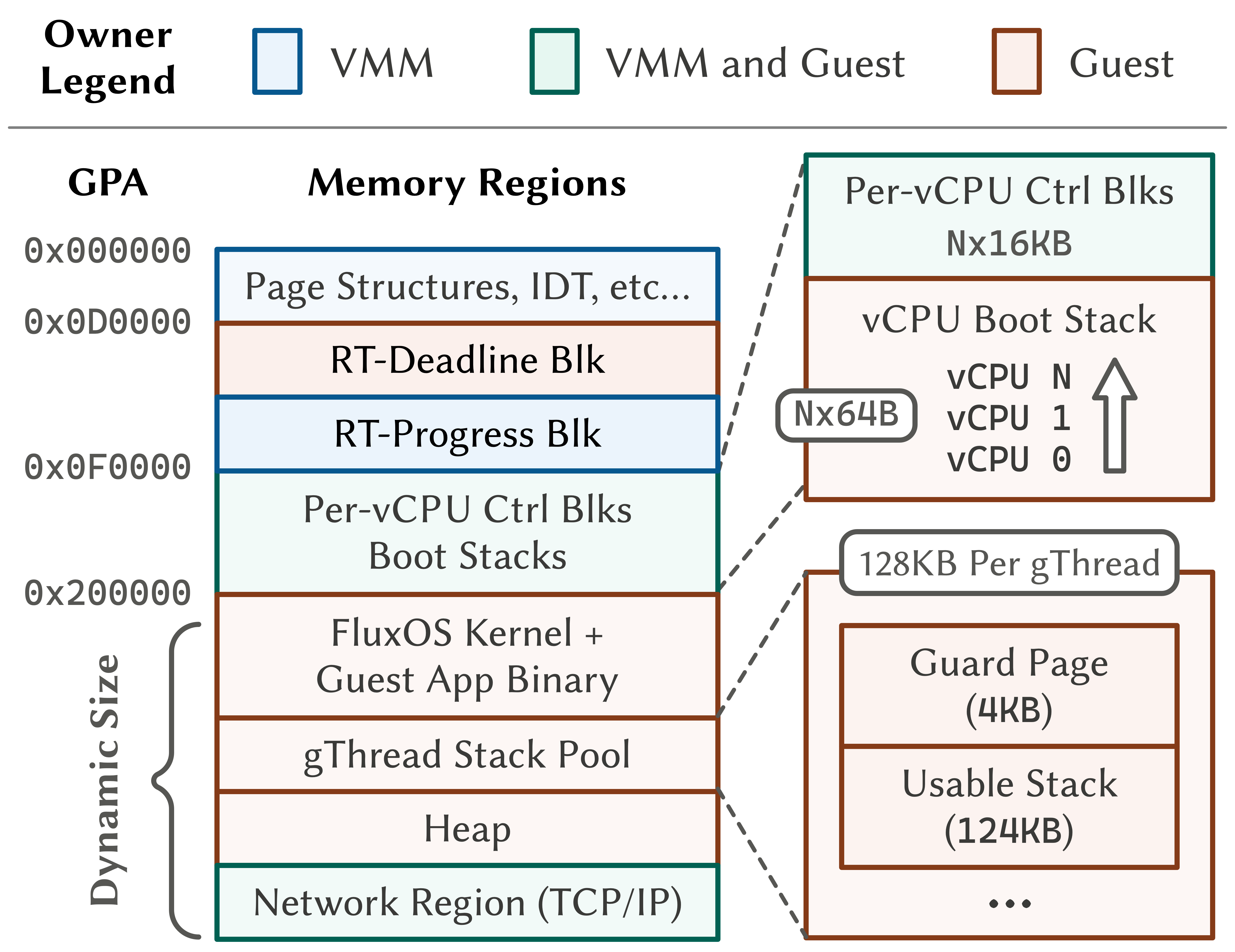}
    \caption{\textsc{FluxOS} and guest application memory layout in a single address space (a simplified view).}
    \label{fig:fluxos-mem}
\end{figure}

\parab{Parallelism and threading.}
Unlike other existing ultralight VMs, \textsc{FluxOS} provides \emph{true multicore} execution. 
\textsc{FluxOS} runs lightweight guest threads (gThreads) across all of the VM's available vCPUs rather than multiplexing them onto one (\secref{sec:missingaxis}).
True multicore is what makes an elastic width worth having, since a VM handed a second core must have runnable work ready to put on it.
A gThread, from the perspective of the guest app, is essentially a user-level thread, similar to a Goroutine or Fiber~\cite{2026go, meta2026folly}.
Concretely, the gThread runs directly in the guest's userspace and is designed to be inexpensive to create;
the runtime cost of creating a gThread is equivalent to that of creating a user-level thread in the host environment, avoiding the conventional heavy guest kernel thread abstraction.
Every active vCPU runs an independent work-stealing scheduler~\cite{blumofe1999scheduling} over a local run-queue and balances load by stealing gThreads from its sibling vCPUs.

\parab{Memory layout.}
\figref{fig:fluxos-mem} shows the guest's physical memory layout.
Beyond the usual paging structures, boot info, and IDT, three regions matter for elasticity: the guest-owned \textit{RT-Deadline} block, where \textsc{FluxOS} publishes its next timer deadline so the VM can downscale to zero vCPUs without spinning one to fire timers; the VMM-owned \textit{RT-Progress} block, where \textsc{Fluxion} signals I/O progress such as arrived packets; and the \textit{Per-vCPU control blocks}, the main \textsc{Fluxion}--\textsc{FluxOS} exchange for demand hints and park/unpark commands, each a tight 64\,B so it does not hurt density.
The \textsc{FluxOS} kernel and guest binary are sized at load time during ELF parsing, and the gThread stack pool and heap follow the configured VM memory size.
% Last, the network region sits at the highest address and is shared with VMM for fast networking.

\parab{Memory footprint.}
\sysname keeps memory footprint small in two ways.
Memory is demand-allocated, so a VM commits only the pages it touches rather than its full configured size.
The page tables are hybrid: \sysname maps the binary and heap with 2\,MB huge pages and falls back to 4\,KB pages only where it must, for the shared control plane and for the gThread stacks, whose guard pages catch overflow.

\subsection{The vCPU Lifecycle}
\label{subsec:lifecycle}

\begin{figure}
    \centering
    \includegraphics[width=\columnwidth]{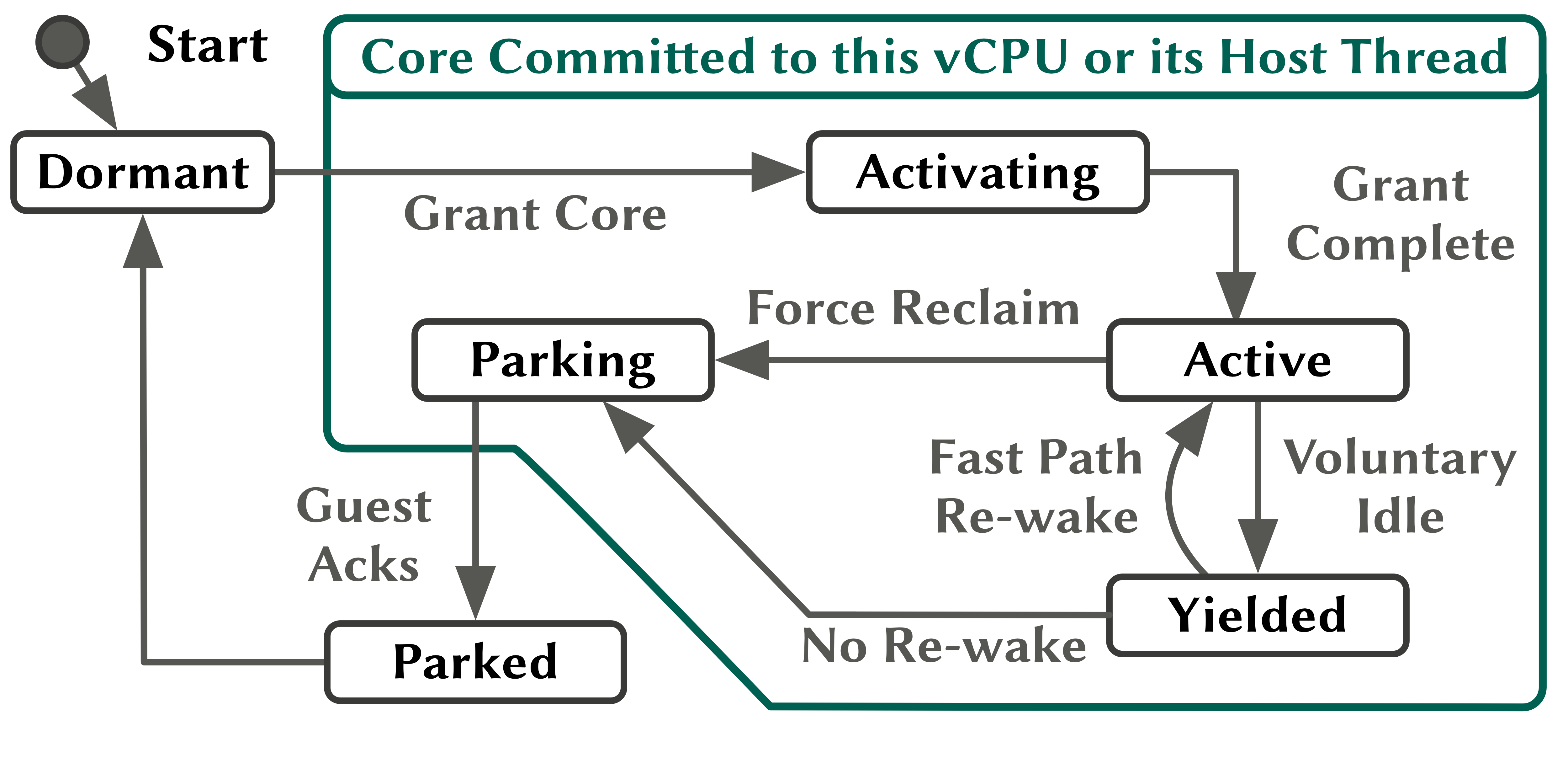}
    \caption{The lifecycle and state transitions of a vCPU.}
    \label{fig:vcpu-lifecycle}
\end{figure}

Every core move in \sysname involves state transitions in a per-vCPU state machine that the host drives and the guest can only request or acknowledge, as illustrated in \figref{fig:vcpu-lifecycle}.
% The subsections that follow trace its edges; we use it here to fix terminology and to pin down what it means for a core to belong to a VM.
A vCPU is in one of six states at any moment of its lifetime.
% partitioned by whether a physical core currently backs it.
In \emph{Activating}, \emph{Active}, \emph{Parking}, and \emph{Yielded}, a physical core is \emph{committed} to the vCPU or its backing host thread: exclusive to it and unavailable to other VMs.
In \emph{Dormant} and \emph{Parked}, the vCPU holds no core and its backing host thread is put back in the queue to be scheduled.
% A Dormant vCPU has none granted; a Parked or Yielded vCPU has released the core it held and sits quiesced, waiting for the host.
% Crossing into the boxed region is how a vCPU acquires a core, and crossing out is how it gives one up.

A vCPU acquires a core when \textsc{Fluxion} grants one, where the grant is staged in \textsf{Activating} and becomes live in \textsf{Active} once \textsc{kFlux} actuates the move (\secref{subsec:decide}).
A vCPU can release a physical core in two situations that differ in who initiates them.
The vCPU's local \textsc{FluxOS} scheduler \emph{voluntarily yields} a physical core when it exhausts its runnable work (with no stealable work either), ceding its core and entering \textsf{Yielded} state, the common case for a bursty tenant between bursts.
A running vCPU can also be \emph{forcibly reclaimed}, which happens when \textsc{Fluxion} decides the core needs to be shifted to another VM (\secref{subsec:reclaim}).
% In this case, \textsc{FluxOS} preempts the current running gThread, puts it back to the run-queue, and marks the run-queue fully stealable.
% This approach allows remainig active vCPUs to help out queued gThreads on parked vCPUs, maximally avoiding work stranding as long as the compute capability allows.
A \textsf{Yielded} vCPU can take the fast re-wake path back to \textsf{Active} if it finds immediately available work to do, \eg a new network packet arrives; this is a performance optimization to avoid relinquishing cores when there are consecutive requests in short intervals.
If there is no immediately available work, the \textsf{Yielded} vCPU transitions into \textsf{Parking} state and eventually relinquishes the physical core.

\subsection{Crossing the Isolation Boundary}
\label{subsec:controlplane}

% Moving a core between VMs requires the host and each guest to agree, continuously and cheaply, on demand and on hand-off.
% Yet a donor and a recipient share neither an address space nor a scheduler, and the shared-memory queues that single-domain runtimes use for this coordination do not survive the KVM/EPT boundary.

\sysname bridges the guest-host boundary with a per-vCPU \emph{control block} mapped into both the guest's physical address space and the host.
Therefore, \textsc{Fluxion} and \textsc{FluxOS} communicate by plain reads and writes to this block: the host writes a one-byte command (such as \textsf{park} or \textsf{activate}) that the guest's per-vCPU scheduler polls, and the guest writes a one-byte status (such as \textsf{park-acked}).
% together with its live demand metrics (\secref{subsec:demand}).
Each field has a single writer, so the block stays coherent under the natural atomicity of aligned word accesses and needs no lock.
On the common path this coordination costs neither a \texttt{VMExit} nor a system call on either side: the host polls 
% (\secref{subsec:decide}) 
and the guest reads the command on its scheduler loop.

When the guest reaches a quiesce point the host is waiting for, such as a completed park or a transition to idle, it must stop the host from spinning on that vCPU; a single doorbell instruction (a port write) forces one VMExit to notify the host, and it is the only trapping primitive on the path.
% Conversely, when the host reads a guest-published value, it must tolerate a writer that is preempted, parked, or hostile mid-update.
% \sysname versions each multi-word publication with a generation counter and reads it with a bounded number of retries, so the host either obtains a consistent snapshot or gives up; it never spins on the guest.
% This second case is where the trust boundary, not just the address-space boundary, shapes the design: the host treats every guest-published byte as a hint it may discard, never as a value it must wait on.

\subsection{Expressing Core Demand}
\label{subsec:demand}

The design of how a \sysname VM can express its core demand is built around two goals:
\begin{inparaenum}[(1)]
    \item \sysname should be able to express and react to core demand change without an explicit core demand indication from the guest app.
    \item \sysname should not rely on the scheduler internals of \textsc{FluxOS}, so that \textsc{FluxOS} may evolve its own scheduling policies without being heavily coupled with the entire infrastructure.
\end{inparaenum}

Greatly inspired by Caladan~\cite{fried2020caladan}, we use the per-vCPU \emph{oldest-ready time} as the primary signal: the timestamp at which the thread now at the head of a vCPU's run-queue became runnable.
The benefit of this type of signal is that it can be directly derived from the \textsc{FluxOS} without any explicit hint from the guest app.
To decouple from \textsc{FluxOS}'s scheduler internals, \textsc{FluxOS} takes the responsibility to publish signals to a narrow communication plane with \textsc{Fluxion}.
Thus, we take several bytes in the per-vCPU control block, serving as the communication plane for core demand signals in a way similar to communicating commands (\secref{subsec:controlplane}).
The main difference between Caladan and \sysname is that \sysname requires \textsc{FluxOS} to continuously publish signals to a designated memory region, while Caladan's control plane directly memory-maps the scheduler and reads values from it without explicit participation from the user application's runtime.
\looseness=-1

The contract is deliberately narrow: \textsc{FluxOS} can retune its scheduler as long as it keeps populating these fields, and \textsc{Fluxion} reads each as a hint, cheap to sample, safe to discard, never a value it must wait on.
\textsc{Fluxion} translates the guest's time-stamp counter to host time using a per-VM offset captured at boot, converts the oldest-ready time to a queueing delay, and takes the maximum across the VM's active vCPUs.
We take the maximum rather than an average because an average dilutes a single thread stuck behind a long-running neighbor with the idle time of the VM's other vCPUs, masking exactly the tail that needs protection.
\textsc{FluxOS} also publishes run-queue depth and the depth of runnable work stranded behind a vCPU that has already parked or yielded, so a VM that is shedding cores still advertises the work it cannot yet run.

% First, \textsc{Fluxion} allocates a single core pool across competing VMs toward a tail-latency objective.
% It therefore needs each VM to expose \emph{how urgently} it needs a core---a pressure signal the policy can rank across tenants---not a core count, which names a solution the allocator cannot reconcile against its neighbors.
% Second, although the host can read guest memory, parsing \textsc{FluxOS}'s scheduler structures directly is the wrong interface: it would couple \textsc{Fluxion} to their private, fast-evolving layout, and those structures mutate concurrently under a lock the host can neither take nor wait on.
% \textsc{FluxOS} therefore reduces its internal state to a small, fixed set of demand signals and publishes them into its control blocks, from which \textsc{Fluxion} derives an allocation trigger.

% A high-priority tenant between bursts may hold no cores at all, yet must reacquire one the instant work arrives.
% \textsc{FluxOS} marks such pending work in the control block, \eg an expiring timer (\secref{subsec:reclaim}).
% \textsc{Fluxion} wakes zero-core VMs on-demand (\eg a timer expired), eliminating the need for these VMs to hold cores to just handle timers.

\parab{Bursty network I/Os and timers.}
An I/O-heavy bursty workload can be idle during bursts and should hold cores only while it is serving requests.
Yet a guest-resident network stack needs timers, and packets arrive when the VM may hold no cores.
\sysname makes both visible to \textsc{Fluxion}.
A gThread blocked on the network parks as \emph{external I/O} and is excluded from the demand signals, so a momentarily idle server still sheds cores instead of advertising phantom demand.
\textsc{FluxOS} also publishes the deadline of its next pending timer in the \textit{RT-Deadline Block}, so \textsc{Fluxion} wakes a zero-core VM exactly when a packet arrives or a timer comes due.
A \sysname VM therefore never holds a core merely to wait, and never misses a wake-up for lack of one, effectively allowing \sysname VMs to downscale to zero cores.
% This mechanism enables \sysname VMs to downscale to zero cores during idle intervals without missing out .

\subsection{Deciding and Actuating a Core Move}
\label{subsec:decide}

\sysname's policy for deciding and enacting a move follows established single-domain core schedulers~\cite{ousterhout2019shenango,fried2020caladan}.
% \secref{sec:related} details what we reuse and what the VM setting changes, and we summarize the mechanism here.
\textsc{Fluxion} runs a busy-polling dataplane pinned to a dedicated core.
Every iteration delivers pending wakes and control responses, while a slow pass, gated to a $10\,\mu$s interval, samples each VM's demand and runs the allocation policy, so policy cost is amortized without delaying wake delivery.
The policy is high-priority-first: \textsc{Fluxion} first hands out cores from a free pool, which takes from no donor, and only then preempts a core from a \emph{structural donor}, a VM running more cores than its guaranteed floor.
It scores candidate recipients by reservation affinity first, then high-priority over low-priority, then the VM furthest below its guaranteed floor;
it admits guaranteed cores as reserved at VM-creation time and damps oscillation with a per-core minimum hold time of $100\,\mu$s, a cooldown on each donor--recipient pair, and a congestion signal cleared as soon as pressure drops.
\sysname arbitrates cores only and does not control shared-cache or SMT-sibling interference.
Actuation is equally lightweight.
\textsc{kFlux}, a host kernel module, facilitates the fast pinning and unpinning of vCPU threads on physical cores.
\textsc{Fluxion} pre-stages the recipient's activation in shared memory before asking the donor to relinquish its core, so the hand-off completes in a single scheduling step once the donor parks.

\subsection{Reclaiming a Core}
\label{subsec:reclaim}

\parab{Reclaiming a core na\"ively strands the guest's work.}
Taking a core back is, in principle, descheduling its vCPU thread from the physical core, but how the host does so decides whether the guest's work might be stranded.
If the host simply lets its scheduler preempt the thread, the core frees at once, yet any guest gThread caught mid-execution is frozen on that vCPU; no other vCPU can resume it, so the work is \emph{stranded} until the same vCPU runs again even when there are other active vCPUs in that VM.
Avoiding this requires coordinating with the guest beforehand, so that in-flight work can be handed to a surviving vCPU later.
We call a gThread \emph{stranded} when it is frozen in a parked vCPU's saved state \emph{while the same VM still has an active vCPU that could run it}: the work could make progress on a sibling core, yet cannot, because no other vCPU can reach into that vCPU's register and stack state.
This avoidable case is what \sysname targets, and it is distinct from a VM that has relinquished all of its cores, whose work is not stranded but suspended together with the VM until the allocator grants it a core again.

\parab{The preemptive-park core hand-off.}
On a forced preemption, \textsc{FluxOS} saves the running gThread's full register state, re-enters its scheduler, marks the gThread as stealable, and acknowledges the park.
As long as the VM retains at least one active vCPU, a sibling vCPU's work-stealing scheduler picks up the marked gThread, so the in-flight work migrates to another of the VM's cores instead of freezing on the parked one.
\sysname therefore guarantees no \emph{avoidable} stranding: whenever a reclaimed VM still holds a core, none of its work is frozen for lack of a runnable vCPU.
When a reclaim takes a VM's last core, the saved gThread stays queued and resumes the moment the VM is next granted a core; this is the defined behavior of a zero-core VM, which the allocator drives to zero only when it has no standing claim to a core (\secref{subsec:decide}), not a stranding failure.
To achieve this hand-off protocol, \textsc{Fluxion} first uses a signal to forcibly trigger a \texttt{VMExit} of the targeted vCPU and return control to the backing host thread.
The backing host thread then injects the interrupt into the vCPU and resumes the vCPU execution, triggering the vCPU's preemption handler immediately upon resuming.

\subsection{Cold Start with Multiple vCPUs}
\label{subsec:coldstart}
\sysname's multi-vCPU machinery adds boot work a single-vCPU ultralight VM does not pay, so keeping cold start fast is an explicit goal.
One of the dominant costs is the core contention among the vCPU-backing host threads, which \textsc{Fluxion} batch-creates under its own core affinity, so a burst of creations piles onto one core and stalls each thread by milliseconds.

\sysname avoids this with a hybrid creation path: 
\begin{inparaenum}[(i)]
    \item It spawns threads locally (inherits \textsc{Fluxion}'s core affinity) for small-vCPU VMs (<8 vCPUs) initially, and then sets the host threads' core affinity to the VM's core set before loading and booting up the full VM.
    \item For large-vCPU VMs, \textsc{Fluxion} delegates the thread creation work to a dedicated helper thread pre-pinned to the target VM-core mask, so threads inherit their placement at birth.
    % a barrier then holds guest execution until all are placed.
\end{inparaenum}

\begin{figure}
    \centering
    \includegraphics[width=0.95\columnwidth]{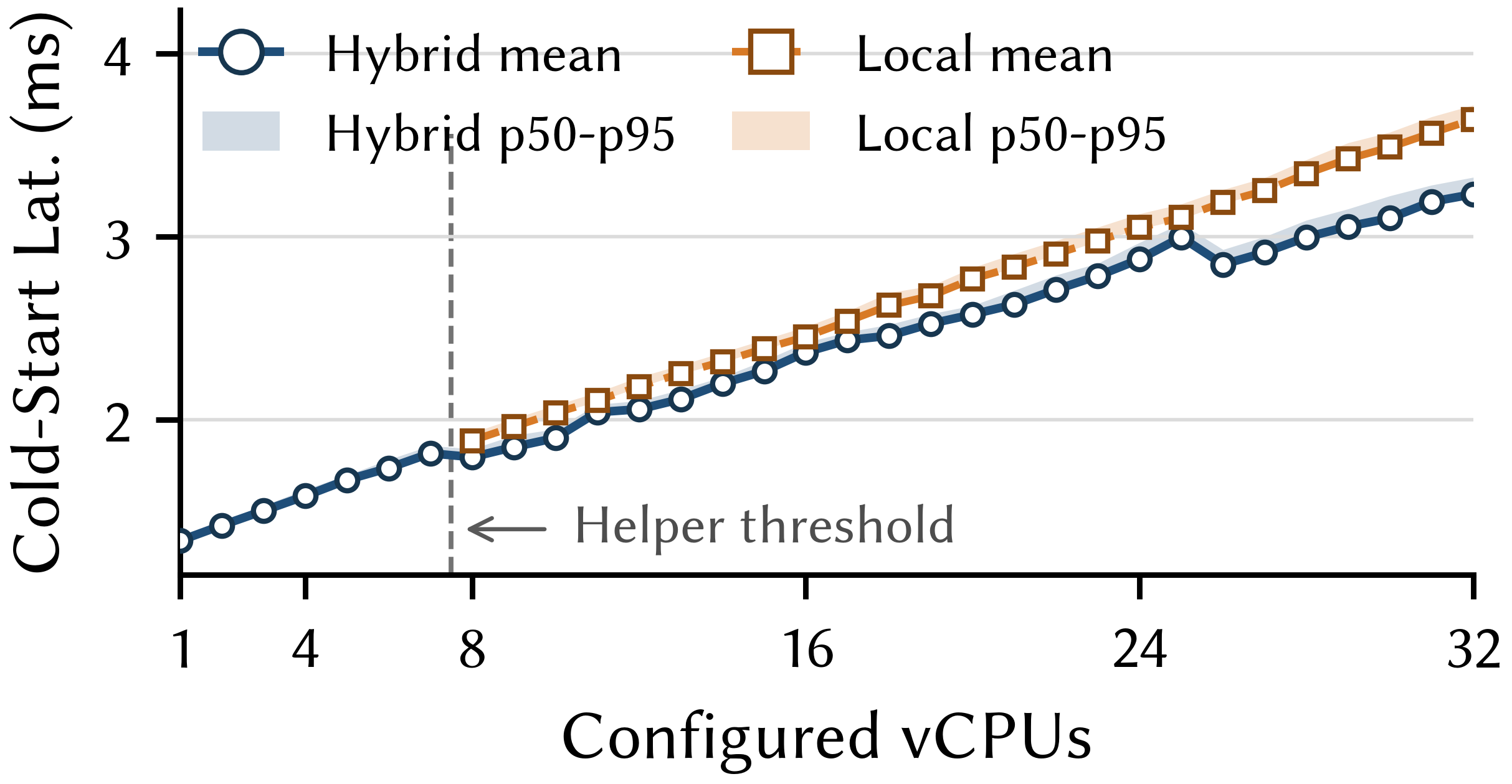}
    \caption{\sysname VM cold-start latency with varying configured vCPUs under a pure local startup policy and the default hybrid startup policy, which uses a helper thread after a threshold.}
    \label{fig:vcpu-cold-start}
\end{figure}

The local spawn saves cross-thread communication overhead for small-scale thread creation, but pays for a short period of core contention.
For large-vCPU VMs, where the core contention is significant, the helper thread approach can avoid the core contention but pays for a small amount of cross-thread communication overhead.
\figref{fig:vcpu-cold-start} shows this cuts a 32-vCPU VM's cold start from 3.64\,ms to 3.23\,ms (11.3\%), with the residual dominated by in-kernel thread creation and the KVM setup any KVM VM pays.

\subsection{The Warm VM Shell}
\label{subsec:warmshell}
To mitigate residual startup latency, \sysname keeps a pool of \textit{warm VM shells}: scaffolds that retain their KVM virtual context (inspired by Virtines~\cite{wanninger2022isolating}), guest memory mapping, and vCPU-backing host threads, so acquiring one skips almost all of the cold path.
\sysname pre-warms shells or reclaims them from terminated VMs (scrubbing guest memory before reuse), and serves a request from the smallest shell whose \textit{warm watermark}, the memory it has already touched, covers the request; it rewrites the boot metadata, control plane, and network state and arms only the requested vCPUs, leaving the rest parked.
We measured warm-shell startup at 100--370\,$\mu$s; this is not a full snapshot restore, as the path still performs ELF parsing and initializes the guest app.

% \subsection{\textsc{FluxOS} Design}
% \subsection{Parallelism in \textsc{FluxOS} (Guest Application)}

% \subsection{Expressing Core Demand}

% \subsection{Voluntary Yielding and Forced Preemption}

% \subsection{Core Allocation and Scheduling}

% \parab{Reclaiming a core na\"ively can strand the guest's work.}
% Taking a core back, in principle, is essentially descheduling its vCPU thread from the physical cores, but how the host does so decides whether the guest's work will be stranded or not.
% If the host simply lets its scheduler preempt the thread, the core frees at once, yet any guest thread caught mid-execution is frozen inside that vCPU's saved register and stack state---no other vCPU can resume it, so the work is \emph{stranded} until the same vCPU runs again.
% Avoiding this requires coordinating with the guest beforehand, so that in-flight work is handed to a surviving vCPU before the core is taken.
% The naive path is fast but unsafe under load; the safe path, when routed through the guest kernel's hot-plug machinery, is slow.
% \sysname's design (\S\ref{sec:design}) turns on closing exactly this gap---preempting and rescuing the in-flight work in microseconds---whereas conventional hot-plug pays a millisecond cost to reconfigure the guest's own view of its CPU set (\S\ref{subsec:designspace}).

% \subsection{Optimizing Cold Start with Multi-vCPUs}
% \yibo{specify the affinity of newly created kernel thread, when }

% \subsection{Warm VM Shell}
% % \yibo{specify the affinity of newly created kernel thread, when }

\subsection{Implementation and Limitations}
\sysname is implemented in Rust and currently supports \texttt{x86\_64} guest applications compiled with \sysname-specified linking procedures.
\sysname has been tested on an \texttt{x86\_64} host server running the Linux kernel and currently only supports KVM.
This restriction is a prototype scope rather than a fundamental one: the core mechanism, parking and unparking vCPU-backing threads with signal-driven \texttt{VMExit} and interrupt injection, has direct equivalents on other KVM-supported architectures such as ARM.
\tabref{tab:loc} summarizes the lines of code for all \sysname components and guest applications ported to \sysname for evaluations.
\textsc{FluxOS} currently does not support any standard Linux syscalls; instead, it supports 24 scheduling/thread primitives and 11 TCP network primitives, offering a sufficiently large surface for guest apps to communicate over the network and to express the workload parallelism.
Beyond networking, \textsc{FluxOS} exposes a read-only host-directory interface for input and reference data: the provider attaches directories at VM creation, and the host resolves each guest request only against them under a strict path grammar that never follows symlinks or names a host path, keeping it within the trust model of \secref{subsec:threat}.
The programming language choice for guest applications is currently limited to Rust, as \textsc{FluxOS} only exposes Rust APIs.
In the serverless model \sysname targets, the provider, not the tenant, selects the guest runtime, so linking against \textsc{FluxOS} is a provider substrate choice rather than a per-tenant porting tax; constrained provider runtimes are already deployed commercially, for instance Cloudflare Workers requires workloads to target its V8 runtime and AWS Lambda ships provider-managed language runtimes.
Supporting unmodified Linux binaries is a known engineering path rather than an open problem: Junction~\cite{fried2024making} already runs unmodified binaries on a Caladan-based runtime, and \textsc{FluxOS}'s scheduler and network stack already follow the same Caladan design, so reaching it requires a fuller ELF loader and a system-call shim that maps the Linux surface onto \textsc{FluxOS} primitives and host services.
% \usepackage{booktabs}
\begin{table}[]
\centering
\resizebox{\columnwidth}{!}{%
\begin{tabular}{@{}lr|lr@{}}
\toprule
\textbf{\sysname Component} & \textbf{LoC} & \textbf{Ported Application} & \textbf{LoC} \\ \midrule
kFlux              & 1,145        & Memcached                   & 2,026        \\
Fluxion            & 13,636       & Webserver                   & 801          \\
FluxOS             & 16,851       & x264                        & 1,200        \\
                   &              & TPC-H                       & 1,147        \\ \bottomrule
\end{tabular}%
}
\caption{Lines of code of \sysname components and ported applications for the purpose of evaluations.}
\label{tab:loc}
\end{table}
\section{Evaluation}
\label{sec:eval}

We evaluate \sysname by answering the following questions:
\begin{enumerate}
    \item Is the \sysname VM lightweight enough compared to existing VM substrates in terms of memory overhead and startup latency, and how fast is its core movement? (\secref{sec:lightweight})
    \item How much does \sysname help the tail latency of high-priority applications during colocation compared to statically sharing cores with existing VM substrates? (\secref{sec:colocation})
    \item Can \sysname effectively redistribute cores to simultaneously protect the high-priority app's tail latency and utilize remaining cores for the low-priority app? (\secref{sec:colocation-distribution})
    \item How does \sysname perform under continuously changing bursty loads compared to using cgroup to adjust core share and hot-plug to adjust core allocation? (\secref{sec:load-change})
\end{enumerate}

These questions evaluate the mechanism that makes dense colocation safe, namely tail protection and demand-driven core redistribution, rather than an end-to-end packing-density or utilization figure, which depends on the colocated tenant mix and provisioning policy and is a deployment-level question.

\parab{Testbed setup:}
We use a single-socket x86 server running Ubuntu~24.04 with KVM, EPT, and VPID enabled.
The server has one Intel\textregistered{} Xeon\textregistered{} Gold~6530 processor (32~cores/64~threads at 2.1\,GHz base), exposed as two NUMA nodes via sub-NUMA clustering, and 256\,GiB of DDR5 memory in total.
We dedicate one NUMA partition to the experiment, isolating its cores with \texttt{isolcpus}, \texttt{nohz\_full}, and \texttt{rcu\_nocbs} for \textsc{Fluxion} and VM vCPUs, while the other partition runs the host OS, background services, and load generators.
We avoid placing competing active vCPUs on SMT siblings for all experiments.
The host is configured for stable microsecond-scale timing: deep c-states are disabled, the clocksource is TSC, and the kernel is booted with \texttt{tsc=reliable}. 
The CPU governor is set to \texttt{performance}, and Intel turbo boost and swap are disabled.

\parab{Evaluated VM substrates and workloads:}
To compare the memory overhead and startup latency, we evaluated Kata~\cite{2026kata} on QEMU and its built-in Dragonball, Unikraft~\cite{kuenzer2021unikraft} on Firecracker, QEMU, and KVM, native Firecracker, Cloud Hypervisor~\cite{2026cloud}, Nanvix~\cite{segarra2026nanvix,penna2026hyperlight}, Dandelion~\cite{kuchler2025unlocking} on KVM, and Hyperlight~\cite{2024hyperlight}.
For experiments that involve colocation, we used four workloads: Memcached~\cite{2026memcached}, x264~\cite{2026corecodec}, custom-built HTTP web server (\textsf{webserver}), and multi-worker TPC-H Q1 query processing with SF-0.1 (\textsf{TPC-H}).
We used native Firecracker (\textsf{FC}) and Cloud Hypervisor (\textsf{CH}) as the baseline VM substrates for colocation experiments.
Firecracker uses the standard virtio-net-over-TAP path and Cloud Hypervisor is evaluated with both TAP (\textsf{CH TAP}) and vhost-user-net (\textsf{CH vhost}) configurations.
For most colocation experiments, we allocated 8 cores for workload VMs. In the four-app-colocation experiment, we allocated 16 cores for workload VMs in total, where each VM is capped at 8 max vCPUs.
Workload VMs can be considered as high-priority (\textsf{HP}) or low-priority (\textsf{LP}):
\textsf{HP} VMs are the ones whose tail latency we want to protect under bursts;
\textsf{LP} VMs are the ones whose cores should ideally be harvested during HP bursts, but that can also harvest cores from HP VMs when HP workloads' core demand is low.
Each workload runs in the form its substrate supports: linked against \textsc{FluxOS} on \sysname, and as an application in a standard Linux guest on Firecracker and Cloud Hypervisor.
Cross-substrate results are therefore system-level comparisons rather than the same binary on both; we isolate the core-movement mechanism itself through the within-\sysname comparison (\secref{sec:colocation-distribution}) and the controlled core-movement microbenchmark (\secref{sec:lightweight}).

\parab{\sysname setup:}
\sysname has two configurations:
\textsf{Hyperflux (auto yield)} automatically relinquishes cores when the scheduler finds no runnable work, and \textsf{Hyperflux (spin)} keeps cores until the host forcibly reclaims them, which represents the lowest possible latency case by avoiding the overhead of yielding and re-waking.
Note that \sysname cannot preempt HP VMs' cores for LP VMs; therefore, when colocating only one HP VM with another LP VM, the spin mode effectively makes the HP VM exclusively use the allocated cores, rendering it incapable of sharing cores with LP VMs.
\looseness=-1

\subsection{Ultralight \sysname VM}
\label{sec:lightweight}
% =========================================================================
% tables/mem-overhead.tex -- host memory footprint of each runtime.
% Include with \input{tables/mem-overhead} and reference as \tabref{tab:mem-overhead}.
%
% Col 2: median resident set size (RSS) of the host-side runtime process while
%        hosting a guest configured with 128 MB of memory.
%        Source: per-target rss_kb_median, converted KiB -> MiB (val / 1024,
%        the same binary base as the 131072 KiB = 128 MB guest config).
% Col 3: footprint relative to \sysname (the normalization baseline),
%        computed from the raw KB medians: rss / 3292.
%        \sysname is therefore the 1.0x reference; <1 is smaller, >1 is larger.
% Rows are sorted by ascending footprint; \sysname is highlighted.
%
% Needs (already in 0_packages.tex): booktabs, colortbl (for \rowcolor).
% lightgreen + \sysname from 0_macros.tex. The \rowcolor guard below keeps the
% table compiling in ACM mode (where xcolor's [table] option may be absent).
% =========================================================================
\providecommand{\rowcolor}[1]{}
\begin{table}[t]
  \centering
  \footnotesize
  \setlength{\tabcolsep}{8pt}
  \renewcommand{\arraystretch}{1.15}
  \begin{tabular}{@{}lrr@{}}
    \toprule
    \textbf{Runtime} & \textbf{RSS (MB)} & \textbf{vs.\ \sysname} \\
    \midrule
    Unikraft (+Firecracker) & 2.5   & $0.8\times$ \\
    \rowcolor{lightgreen}
    \textbf{\sysname}        & \textbf{3.2} & \textbf{$1.0\times$} \\
    Hyperlight               & 3.6   & $1.1\times$ \\
    Dandelion (+KVM)                & 6.0  & $1.9\times$ \\
    Nanvix                   & 20.8  & $6.5\times$ \\
    Unikraft (+QEMU)        & 37.1  & $11.5\times$ \\
    Unikraft (+KVM)         & 38.4  & $11.9\times$ \\
    Firecracker              & 50.5  & $15.7\times$ \\
    Cloud Hypervisor         & 52.9  & $16.5\times$ \\
    Kata (+Dragonball)      & 170.3 & $53\times$ \\
    Kata (+QEMU)            & 337.7 & $105\times$ \\
    \bottomrule
  \end{tabular}
  \caption{Host memory footprint (resident set size, RSS) of each runtime while hosting a guest configured with 128\,MB of memory. 
  The last column reports each footprint relative to \sysname{}.}
  \label{tab:mem-overhead}
\end{table}

To compare the memory overhead and startup latency, we used a simple ``Hello World'' application ported to each substrate so that we can focus on the substrate itself.

\parab{Memory overhead.}
\tabref{tab:mem-overhead} shows that the memory overhead of \sysname (3.2\,MB) is comparable to existing ultralight VM substrates such as Hyperlight, Dandelion, and Nanvix, and, as expected, significantly lower than that of microVM-league substrates, such as Cloud Hypervisor and Firecracker.
Unikraft on Firecracker has the lowest memory overhead and wins over \sysname by just 0.7\,MB.

\begin{figure}
    \centering
    \includegraphics[width=\columnwidth]{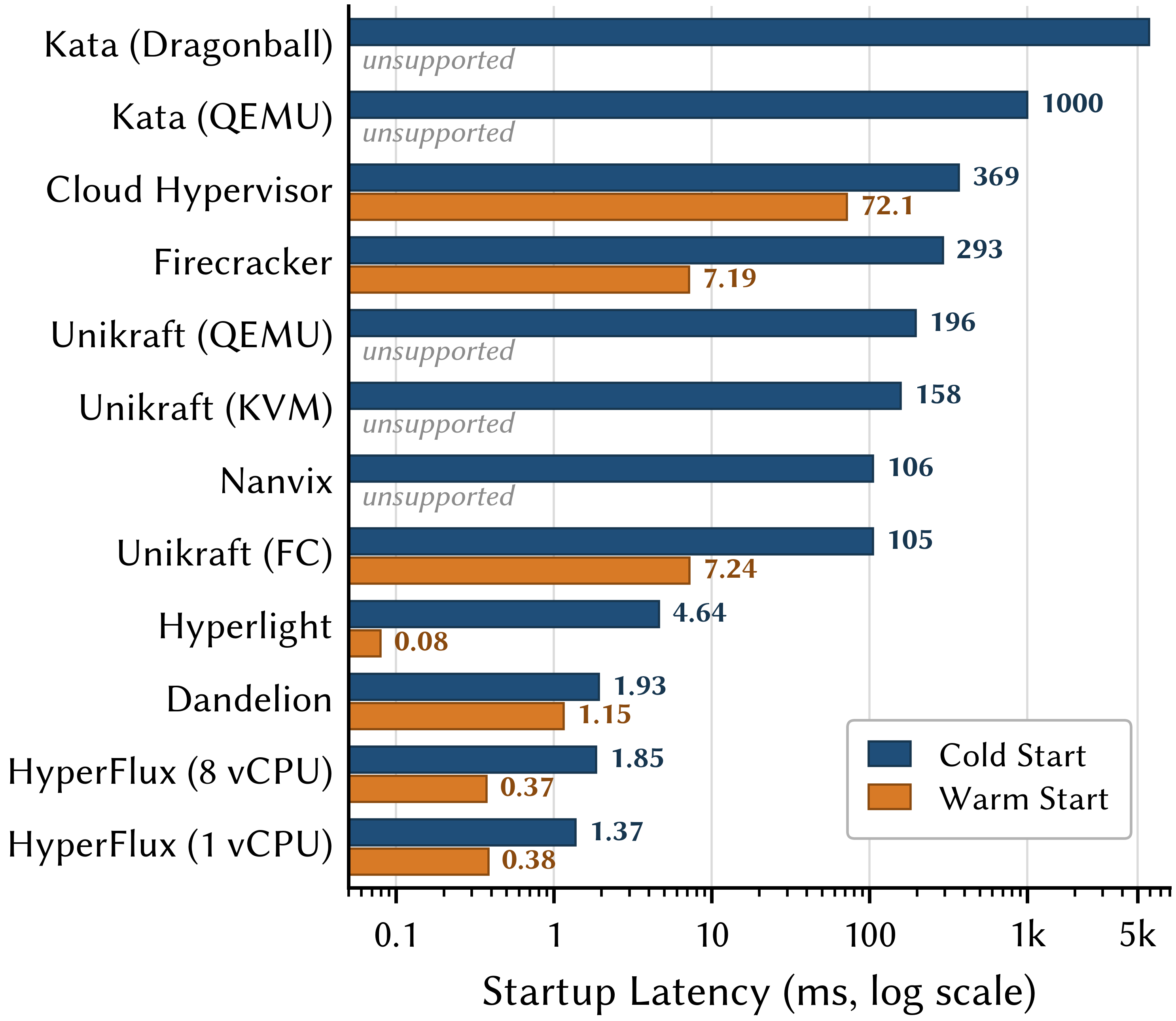}
    \caption{Startup latency comparison. \textbf{Note}: \sysname currently does not support a full instance snapshot, so \sysname warm start only uses a pre-warmed VM shell. \sysname still needs to perform ELF parsing, load the guest binary, and initialize the guest app.}
    \label{fig:startup-latency}
\end{figure}

\parab{Startup latency.}
\figref{fig:startup-latency} shows both cold-start latency and warm-start latency (if supported) among all evaluated substrates.
\sysname's cold-start latency (1.37\,ms with 1 vCPU and 1.85\,ms with 8 vCPUs) is on par with the fastest existing ultralight VM substrates.
However, the \sysname VM supports multi-core parallelism with threading while the other ultralight VM substrates do not.
\sysname's warm start with a warm VM shell has a higher latency than Hyperlight (0.38\,ms vs. 0.11\,ms); however, we note that \sysname's number is not fully comparable to Hyperlight's as \sysname currently does not support a full instance snapshot, and \sysname still needs to perform ELF parsing and initialize guest app memory during the warm-start measurement.

\begin{table}[]
\centering
\resizebox{\columnwidth}{!}{%
\begin{tabular}{@{}lrrrr@{}}
\toprule
\textbf{Operation} & \textbf{p50} & \textbf{p99} & \textbf{max} & \textbf{mean} \\ \midrule
Remove (reclaim core from busy donor) & 7.4 & 9.0  & 10.3 & 7.4 \\
Add (install core on recipient)       & 4.4 & 25.8 & 33.4 & 5.6 \\ \bottomrule
\end{tabular}%
}
\caption{\sysname forced cross-VM core-movement latency ($\mu$s, $n{=}1000$): a busy donor vCPU is forcibly preempted, timed from the preemption signal until its core is released ({Remove}) and re-installed on a recipient ({Add}).}
\label{tab:hyperflux-core-realloc}
\end{table}

\parab{Core movement latency.}
\tabref{tab:hyperflux-core-realloc} reports the latency of a forced cross-VM core movement, measured with fenced host-TSC reads on an otherwise idle host.
To exercise the worst case rather than a cooperative yield, we forcibly preempt a \emph{spinning}, non-yielding donor vCPU: \sysname signals it out of guest execution and injects the preemption interrupt, \textsc{FluxOS} saves the running gThread and acknowledges the park, and we time from the preemption signal until the host core is released and reassignable to another VM.
Across 1000 forced moves, reclaiming a core from a busy donor takes 7.4\,$\mu$s on average (p99 9.0\,$\mu$s) and installing it on a recipient takes 5.6\,$\mu$s on average (median 4.4\,$\mu$s, p99 25.8\,$\mu$s), so a full cross-VM harvest averages $\sim$13\,$\mu$s, orders of magnitude faster than the ACPI vCPU hot-plug path in QEMU and Cloud Hypervisor (\textasciitilde100--200\,ms).
The reclaim path is tightly bounded because the host drives it with a preemption signal, whereas the wider tail on the install path comes from waking the parked vCPU thread and re-entering the guest on the host side, not from guest cooperation, and even its p99 stays well below the millisecond-scale hot-plug path.
These figures are the actuation cost of the move itself; reacting to a burst additionally incurs \textsc{Fluxion}'s demand sampling and policy pass (\secref{subsec:decide}).
The measurement excludes the benchmark's external control-socket round trip, since in deployment a move is triggered by \textsc{Fluxion}'s own dataplane loop rather than an external call.

\begin{figure}
    \centering
    \includegraphics[width=\columnwidth]{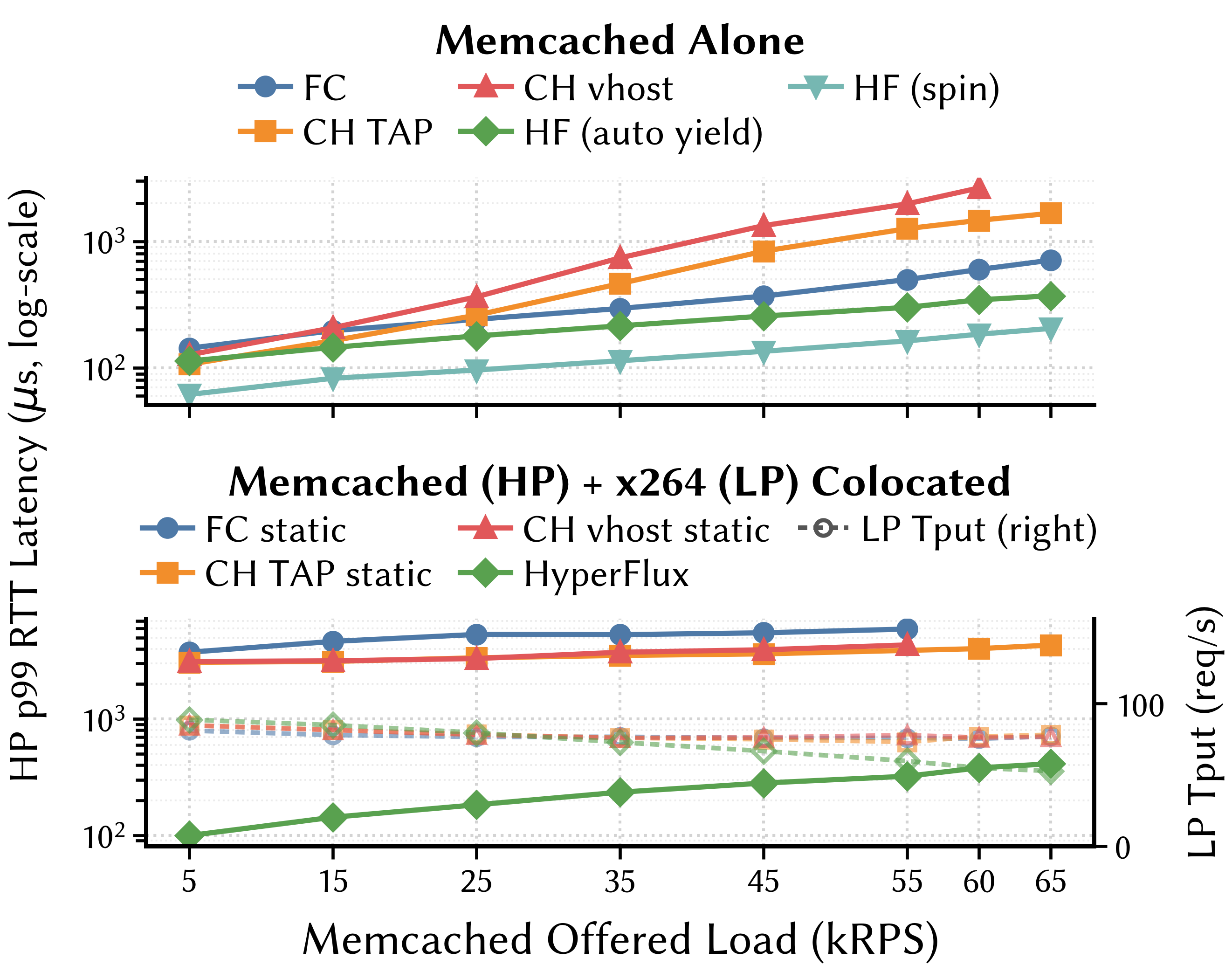}
    \caption{
        \sysname shows a lower latency compared to Firecracker (\textsf{FC}) and Cloud Hypervisor with different network backends (\textsf{CH vhost/TAP}) in both isolation (top) and colocation (bottom).
        Label \textsf{static} means there are no dynamic core-share changes between colocated applications.
        Missing data points indicate the corresponding setting is overloaded under its offered load.
        }
    \label{fig:multi-runtime}
\end{figure}

\subsection{Colocation with Static Core Allocation}
\label{sec:colocation}

\begin{figure*}
    \centering
    \includegraphics[width=\textwidth]{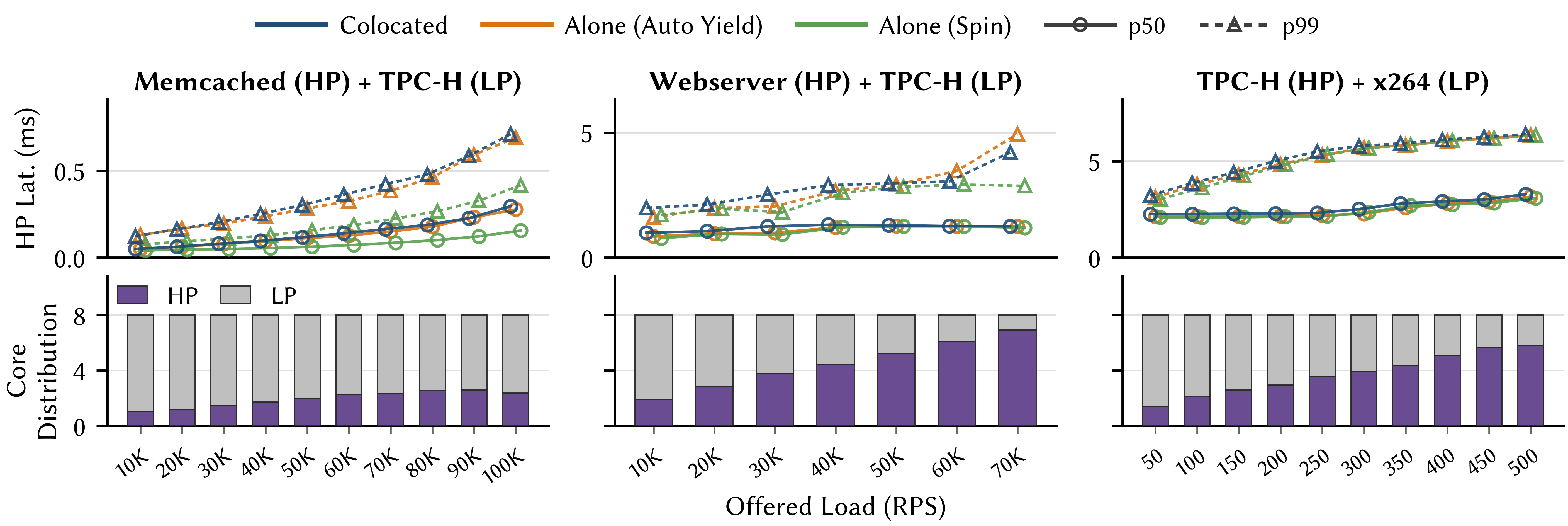}
    \caption{\sysname colocation shows a close latency measurement compared to running the high-priority (\textsf{HP}) application alone with automatic core yielding (\textsf{Alone (Auto Yield)}). Memcached and TPC-H pairing shows that automatic core yielding has runtime overhead compared to the spinning case (\textsf{Alone (Spin)}), where the Memcached VM spins on cores without releasing them during idle intervals.}
    \label{fig:mini-colocation}
\end{figure*}

We further evaluate \sysname's effectiveness in protecting the tail latency of the HP workload VM by comparing it with statically sharing cores with Firecracker and Cloud Hypervisor respectively.

We first measured the p99 RTT latency of Memcached under different offered loads when it is running alone to establish the baseline.
\figref{fig:multi-runtime} (top) shows that \sysname (auto yield) delivers consistently lower latency than Firecracker and Cloud Hypervisor, regardless of the network backend (\textsf{CH TAP/vhost}).
All configurations have a similar p99 latency at low load (10K), but the gap widens as the load increases.
\sysname performs better even when the workload runs alone because \sysname can wake up targeted vCPUs to immediately handle the pending network packet.
\figref{fig:multi-runtime} (top) also reveals that \sysname (spin) has even lower latency compared to \sysname (auto yield) as it keeps spinning on idle cores rather than relinquishing them, waiting for the next ready packet to process. 
Therefore, \sysname (spin) avoids the virtual machine context switch cost.
Here, with no competing HP VM, \sysname (spin) effectively gives the HP VM exclusive use of its cores, which the colocated LP VM cannot harvest.

\figref{fig:multi-runtime} (bottom) shows the latency under static colocation, where both VMs share the same eight cores and the host OS scheduler multiplexes their vCPU threads.
In this case, \sysname holds a substantial p99 latency advantage over both Firecracker and Cloud Hypervisor under static colocation (up to 10.5x lower at 65K load).
This gap reflects two compounding effects: \sysname's runtime already leads when the HP VM runs alone (\figref{fig:multi-runtime}, top), and demand-aware core movement then preserves that lead under contention, an effect isolated by the within-\sysname comparison in \secref{sec:colocation-distribution}.
Dashed lines plot the LP workloads' throughput and show that static colocations have mostly flat throughput lines while \sysname gracefully degrades LP workload throughput under high load of Memcached's traffic.
% This result validates that \sysname can throttle the LP VM to protect the HP workload's latency much better than the static colocation approaches.

\begin{figure}
    \centering
    \includegraphics[width=\columnwidth]{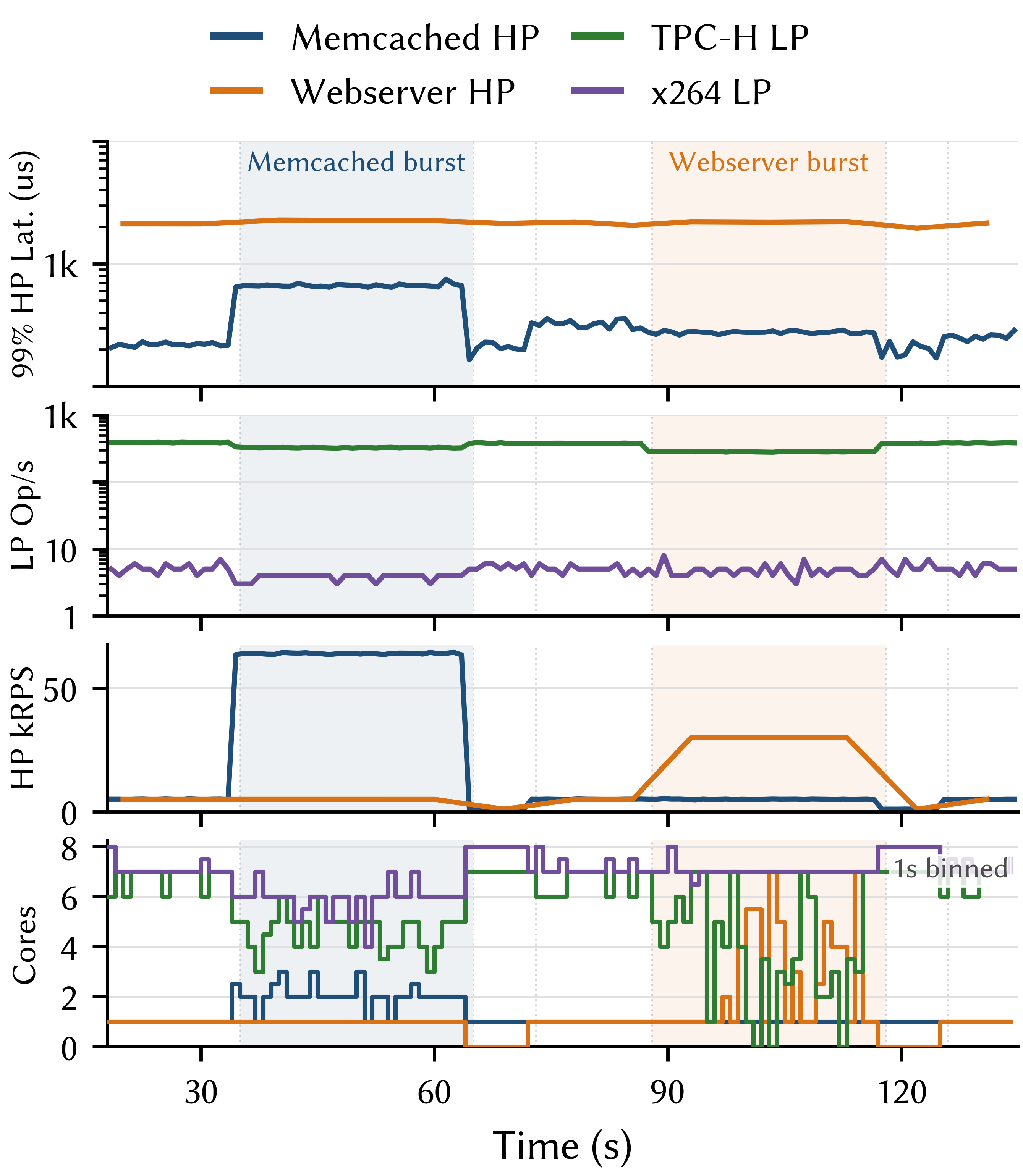}
    \caption{A four-app-colocation scenario, where \sysname can actively and promptly shift cores from low-priority (\textsf{LP}) apps to high-priority (\textsf{HP}) apps when HP apps are under bursty loads.}
    \label{fig:colocation-2x2}
\end{figure}

% \parab{Latency increases under higher loads.}
% The higher Memcached p99 we report (>100\,$\mu$s vs. $\sim$80\,$\mu$s in a conventional setup) comes mainly from queueing delay in both the host network stack (our load generators send from the host into the VM) and the guest network stack.

\subsection{Core Distribution among Colocated Workloads}
\label{sec:colocation-distribution}

\parab{Colocation with two workload VMs.}
To understand how \sysname distributes cores among colocated workload VMs, we evaluate three HP-LP pairings: Memcached-TPC-H, Webserver-TPC-H, and TPC-H-x264.
\figref{fig:mini-colocation} (top) shows that under \sysname the HP VM's colocated tail latency stays close to its alone latency across all three pairings.
Because both runs use the same \sysname runtime, this within-\sysname comparison controls for the substrate's runtime advantage: the small alone-to-colocated gap reflects what elastic core movement buys under colocation, not a head start from the lightweight runtime.
In the Memcached-TPC-H pairing, the colocated Memcached tail is higher than in the alone case with spinning, as expected, because spin mode occupies all cores with busy polling for the next packet.

As shown in \figref{fig:mini-colocation} (bottom), in both Webserver-TPC-H and TPC-H-x264 pairings, \sysname demonstrates effective core distribution in two ways:
it allocates more cores to the LP VM when the HP VM's load is low and it has free cores to share;
it also harvests more cores from the LP VM when the HP VM's load is high and it needs more cores.
That the colocated HP tail stays near its alone-case (top) is the visible result of these cores following demand to the HP VM, while the LP VM still receives cores between bursts, not of the LP VM simply failing to interfere.
The Memcached-TPC-H pairing shows the HP VM does not gain substantially more cores under higher loads.
This pattern arises because Memcached's server-side service time is only tens of microseconds, so demand fluctuates faster than \sysname reacts, and its parallelism is not aggressive enough to exploit the extra cores \sysname can offer.
\looseness=-1
% We note that any VM can be set as either an HP or LP VM, and, as we have demonstrated, TPC-H can be configured as either.
% \sysname can treat the workload VM accordingly and effectively prioritize the core usage for the HP VM as long as it can exploit the parallelism that \sysname offers.

\parab{Colocation with many workload VMs.}
To further understand \sysname's core distribution behavior in a more complex scenario, we evaluate \sysname with a four-app-colocation scenario, where two HP VMs (Memcached and Webserver) are colocated with two LP VMs (TPC-H and x264).
\figref{fig:colocation-2x2} shows the HP workloads' p99 latencies, LP workloads' throughputs, offered HP loads, and core distribution over a $\sim$120\,s timespan.
During both Memcached and Webserver bursts, \sysname harvests cores from both LP VMs and shifts them to the HP VMs, which effectively keeps the HP VMs' latency low and stable.
When the HP VMs are not under bursty load, \sysname also gives more cores to the LP VMs so that they can achieve higher throughput.
The rise of p99 latency during the Memcached burst is expected due to the increased queueing delay in the host network stack.
% as we have observed in \secref{sec:colocation}.
During the idle period when the offered Webserver load is zero (right after the burst), \sysname also gracefully parks all of the Webserver VM's cores, giving up core resources so that other VMs can harvest.

\subsection{Effectiveness of \sysname under Continuously Changing Bursty Loads}
\label{sec:load-change}

Next, we evaluate whether \sysname can move cores promptly and effectively under continuously changing loads, comparing it with two existing approaches:
using cgroup to adjust the core share and using vCPU hot-plug to adjust the core allocation among colocated workload VMs.
We colocate Memcached (HP) and x264 (LP) together and change the offered load to Memcached every 5 seconds in a three-phase pattern (10K, 30K, and 50K).

\begin{figure}
    \centering
    \begin{subfigure}{\columnwidth}
        \centering
        \includegraphics[width=\columnwidth]{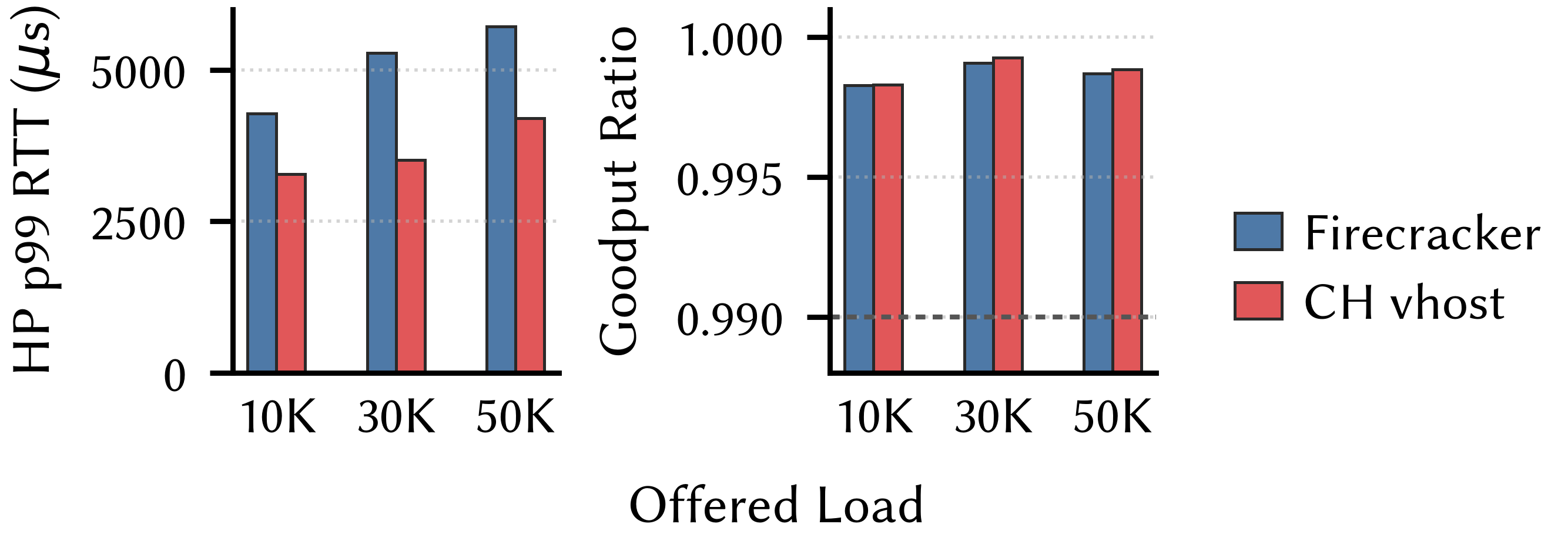}
        \caption{
            Colocation by statically sharing 8 cores under three steady loads.
            % which establishes the baseline for the bursty phased colocation scenario. 
        Goodput ratio (right) suggests both \textsf{FC} and \textsf{CH vhost} can sufficiently handle the offered load (>99.5\% of the offered load) under a steady traffic flow.}
        \label{fig:static-gput}
    \end{subfigure}

    \begin{subfigure}{\columnwidth}
        \centering
        \includegraphics[width=\columnwidth]{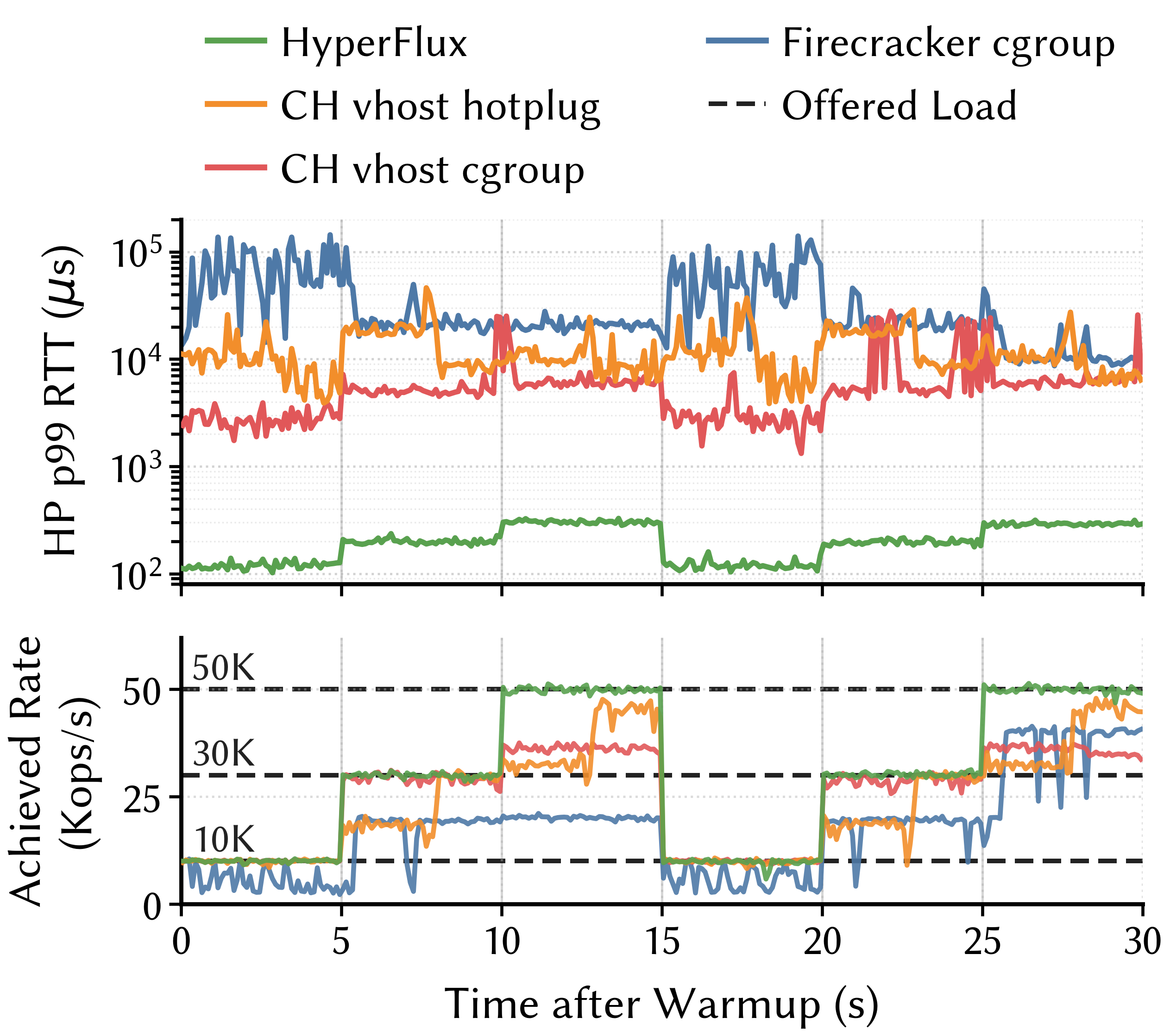}
        \caption{Colocation under a three-phase bursty traffic pattern (10K, 30K, and 50K), where \textsf{Firecracker} uses cgroup to adjust the core share dynamically and \textsf{CH vhost} additionally uses hot-plug to adjust the core allocation. 
        \sysname can promptly shift cores during Memcached bursts by showing an immediate change in the achieved rate (bottom) with a lower and more stable p99 latency (top). 
        Hot-plug shows a delayed response and cgroup shows an ineffective core share under the 50K load.}
        \label{fig:bursty-phased}
    \end{subfigure}
    \caption{Memcached (HP) and x264 (LP) colocation.}
    \label{fig:colocation-memcached}
\end{figure}

As a baseline, \figref{fig:static-gput} shows that static colocation (\secref{sec:colocation}) with both Firecracker and Cloud Hypervisor serves all three steady loads at >99.5\% goodput (achieved/offered).
Then, we proceed to experiment with Firecracker and Cloud Hypervisor using cgroup and hot-plug, and \sysname, under the three-phase load.
We use cgroup to adjust the time quota, which gives a 2:6 share to the Memcached VM and the x264 VM under the 10K load, a 4:4 share under the 30K load, and a 6:2 share under the 50K load.
We use hot-plug to adjust the core allocation between the Memcached VM and the x264 VM, offering 2 and 6 cores to the two VMs under the 10K load, 4 and 4 cores under the 30K load, and 6 and 2 cores under the 50K load.
Both actuators are oracle-based: they cannot detect load changes but react exactly when the load shifts, using the phase schedule as prior knowledge. This represents the best possible scenario for cgroup and hot-plug.
In contrast, \sysname reacts based on the actual runtime demand and performs core movement dynamically, thus incurring reaction overhead.

\figref{fig:bursty-phased} depicts the result. As shown in the HP p99 latency (top), \sysname consistently maintains a lower and more stable tail latency while other systems fluctuate significantly.
The achieved rate (bottom) shows two main insights:
\begin{inparaenum}[(1)]
   \item Based on the \textsf{CH vhost} result, the cgroup approach is effective at shifting cores under load changes but is insufficient at the high load (50K), where it fails to deliver the full 50K traffic load.
   This reveals the fundamental limitation of the time-share mechanism. 
   Under such a high load, exclusive access to a core is more crucial to protect the tail latency.
   \item The \textsf{CH vhost hotplug} shows that the vCPU hot-plug approach is too slow to react to drastic load changes.
   For instance, at the 5s and 10s time-points, Cloud Hypervisor with vCPU hot-plug takes about 2\,s to reach the intended rate.
   This far exceeds a single ACPI hot-plug core movement (\textasciitilde100--200\,ms) for two compounding reasons: each phase onlines several guest CPUs in sequence, and while those cores are still being added the under-provisioned VM accumulates a request backlog that it must drain before its achieved rate catches up to the offered load.
%    Slow actuation thus pays a double penalty, the hot-plug latency plus the backlog that accrues while the VM waits, whereas \sysname adds capacity in microseconds, before a backlog can form.
\end{inparaenum}
In summary, \sysname can effectively deliver the offered load immediately after the load change and consistently satisfy offered loads across three phases.
\looseness=-1

\section{Related Work}
\label{sec:related}

\parab{Microsecond-scale core scheduling.}
Caladan, Shenango, and Arachne reallocate cores among single-trust-domain applications in microseconds, and Shinjuku, ZygOS, and Nu schedule requests and resources just as fast~\cite{fried2020caladan, ousterhout2019shenango, qin2018arachne, kaffesshinjukupreemptivescheduling2019, prekaszygosachievinglow2017, ruan2023nu}, showing the timescale is achievable on commodity hardware.
\sysname's core arbiter and kernel actuation path adapt the substrate they pioneered~\cite{fried2020caladan, ousterhout2019shenango}; what is new is carrying it across a VM isolation boundary without sacrificing ultralight-VM startup speed and footprint.
\sysname is to colocated VMs what these schedulers are to colocated application processes.

\parab{Cross-VM core harvesting.}
Azure's resource-harvesting VMs, SmartHarvest, and memory-harvesting VMs reclaim a primary VM's spare capacity for batch tenants, but only asymmetrically and when idle~\cite{ambati2020providing, wang2021smartharvest, fuerstmemoryharvestingvmscloud2022}; HardHarvest reaches microseconds only with new hardware~\cite{stojkovic2025hardharvest}; and Ditto assumes a strict parallelism paradigm~\cite{zhao2024ditto}.
Each thus leaves one of \sysname's axes uncovered (\secref{sec:missingaxis}, \tabref{tab:position}).
% None reclaim \emph{running} work from a non-yielding peer and redistribute it among equals at microsecond scale on commodity KVM.

\parab{vCPU scheduling and resizing.}
vScale and FlexCore scale an SMP guest's active vCPUs via ballooning~\cite{cheng2016vscale, miaoflexcoredynamicvirtual2015}, and other work coordinates guest and host scheduling to avoid lock-holder preemption and double scheduling~\cite{uhlig2004scalable, kim2013demandbased}.
All act within one guest's allotment through its kernel; none hand a physical core to a \emph{different} VM.
\sysname instead makes the physical core, not the vCPU count, the unit of allocation and moves cores across VMs in microseconds.
% : the count stays fixed while the host parks and re-backs vCPU threads to move cores across guests in microseconds (\secref{sec:design}).

\parab{Lightweight virtualization and software isolation.}
\sysname inherits the lightweight-VM lineage that trades functionality for agility~\cite{agachefirecrackerlightweightvirtualization2020, 2026cloud, 2026kata, madhavapeddyunikernelsrisevirtual2013, kuenzer2021unikraft, kuchler2025unlocking, penna2026hyperlight, wanninger2022isolating}, including cold-start work like Catalyzer and RunD~\cite{du2020catalyzer, li2022rund}, preserving their fast boot speed and low memory footprint, while also keeping hardware-based isolation and supporting elastic parallelism width.
The densest forgo the in-VM multicore execution; Nanvix restores threading atop Hyperlight, but on a single vCPU~\cite{segarra2026nanvix}.
Others instead weaken or relocate the isolation boundary for density with software isolation or container-based approaches, \eg WebAssembly~\cite{haasbringingwebspeed2017}, Faasm~\cite{shillakerfaasmlightweightisolation2020}, GRANNY~\cite{segarra2025granny}, gVisor~\cite{2026gvisor}, and Junction~\cite{fried2024making}.

\section{Conclusion}
\label{sec:conclusion}

Serving bursty, latency-sensitive serverless tenants at high density demands three properties at once: a small footprint and fast cold start, in-VM multicore execution, and microsecond movement of physical cores across the VM isolation boundary.
No prior substrate offered all three; \sysname is the first to do so on commodity KVM.
By parking and re-backing vCPU threads instead of reconfiguring the guest's CPU topology, it moves a core between colocated VMs in microseconds, orders of magnitude faster than the ACPI hot-plug path it supplants.
That speed turns cross-VM core movement into a viable tail-latency mechanism rather than a coarse provisioning tool: \sysname steers cores to high-priority tenants under load and returns them between bursts while staying on par with the densest ultralight VMs, making elastic parallelism width, a previously missing axis in the VM design space, attainable on commodity KVM.

\newpage
%% ============================================================
%% Bibliography (explicit commands are visible to arXiv's source scanner)
%% ============================================================
\ifneurips
  \bibliographystyle{plainnat}
\else\ifusenix
  \bibliographystyle{plain}
\else
  \bibliographystyle{templates/acmart/ACM-Reference-Format-No-Year}
\fi\fi
\bibliography{hyperflux}

\newpage
\appendix

\ifneurips
  \newpage
  \section*{NeurIPS Paper Checklist}

%%% BEGIN INSTRUCTIONS %%%
The checklist is designed to encourage best practices for responsible machine learning research, addressing issues of reproducibility, transparency, research ethics, and societal impact. Do not remove the checklist: {\bf The papers not including the checklist will be desk rejected.} The checklist should follow the references and follow the (optional) supplemental material.  The checklist does NOT count towards the page
limit. 

Please read the checklist guidelines carefully for information on how to answer these questions. For each question in the checklist:
\begin{itemize}
    \item You should answer \answerYes{}, \answerNo{}, or \answerNA{}.
    \item \answerNA{} means either that the question is Not Applicable for that particular paper or the relevant information is Not Available.
    \item Please provide a short (1--2 sentence) justification right after your answer (even for \answerNA). 
   % \item {\bf The papers not including the checklist will be desk rejected.}
\end{itemize}

{\bf The checklist answers are an integral part of your paper submission.} They are visible to the reviewers, area chairs, senior area chairs, and ethics reviewers. You will also be asked to include it (after eventual revisions) with the final version of your paper, and its final version will be published with the paper.

The reviewers of your paper will be asked to use the checklist as one of the factors in their evaluation. While \answerYes{} is generally preferable to \answerNo{}, it is perfectly acceptable to answer \answerNo{} provided a proper justification is given (e.g., error bars are not reported because it would be too computationally expensive'' or ``we were unable to find the license for the dataset we used''). In general, answering \answerNo{} or \answerNA{} is not grounds for rejection. While the questions are phrased in a binary way, we acknowledge that the true answer is often more nuanced, so please just use your best judgment and write a justification to elaborate. All supporting evidence can appear either in the main paper or the supplemental material, provided in appendix. If you answer \answerYes{} to a question, in the justification please point to the section(s) where related material for the question can be found.

IMPORTANT, please:
\begin{itemize}
    \item {\bf Delete this instruction block, but keep the section heading ``NeurIPS Paper Checklist"},
    \item  {\bf Keep the checklist subsection headings, questions/answers and guidelines below.}
    \item {\bf Do not modify the questions and only use the provided macros for your answers}.
\end{itemize}

%%% END INSTRUCTIONS %%%

\begin{enumerate}

\item {\bf Claims}
    \item[] Question: Do the main claims made in the abstract and introduction accurately reflect the paper's contributions and scope?
    \item[] Answer: \answerTODO{} % Replace by \answerYes{}, \answerNo{}, or \answerNA{}.
    \item[] Justification: \justificationTODO{}
    \item[] Guidelines:
    \begin{itemize}
        \item The answer \answerNA{} means that the abstract and introduction do not include the claims made in the paper.
        \item The abstract and/or introduction should clearly state the claims made, including the contributions made in the paper and important assumptions and limitations. A \answerNo{} or \answerNA{} answer to this question will not be perceived well by the reviewers. 
        \item The claims made should match theoretical and experimental results, and reflect how much the results can be expected to generalize to other settings. 
        \item It is fine to include aspirational goals as motivation as long as it is clear that these goals are not attained by the paper. 
    \end{itemize}

\item {\bf Limitations}
    \item[] Question: Does the paper discuss the limitations of the work performed by the authors?
    \item[] Answer: \answerTODO{} % Replace by \answerYes{}, \answerNo{}, or \answerNA{}.
    \item[] Justification: \justificationTODO{}
    \item[] Guidelines:
    \begin{itemize}
        \item The answer \answerNA{} means that the paper has no limitation while the answer \answerNo{} means that the paper has limitations, but those are not discussed in the paper. 
        \item The authors are encouraged to create a separate ``Limitations'' section in their paper.
        \item The paper should point out any strong assumptions and how robust the results are to violations of these assumptions (e.g., independence assumptions, noiseless settings, model well-specification, asymptotic approximations only holding locally). The authors should reflect on how these assumptions might be violated in practice and what the implications would be.
        \item The authors should reflect on the scope of the claims made, e.g., if the approach was only tested on a few datasets or with a few runs. In general, empirical results often depend on implicit assumptions, which should be articulated.
        \item The authors should reflect on the factors that influence the performance of the approach. For example, a facial recognition algorithm may perform poorly when image resolution is low or images are taken in low lighting. Or a speech-to-text system might not be used reliably to provide closed captions for online lectures because it fails to handle technical jargon.
        \item The authors should discuss the computational efficiency of the proposed algorithms and how they scale with dataset size.
        \item If applicable, the authors should discuss possible limitations of their approach to address problems of privacy and fairness.
        \item While the authors might fear that complete honesty about limitations might be used by reviewers as grounds for rejection, a worse outcome might be that reviewers discover limitations that aren't acknowledged in the paper. The authors should use their best judgment and recognize that individual actions in favor of transparency play an important role in developing norms that preserve the integrity of the community. Reviewers will be specifically instructed to not penalize honesty concerning limitations.
    \end{itemize}

\item {\bf Theory assumptions and proofs}
    \item[] Question: For each theoretical result, does the paper provide the full set of assumptions and a complete (and correct) proof?
    \item[] Answer: \answerTODO{} % Replace by \answerYes{}, \answerNo{}, or \answerNA{}.
    \item[] Justification: \justificationTODO{}
    \item[] Guidelines:
    \begin{itemize}
        \item The answer \answerNA{} means that the paper does not include theoretical results. 
        \item All the theorems, formulas, and proofs in the paper should be numbered and cross-referenced.
        \item All assumptions should be clearly stated or referenced in the statement of any theorems.
        \item The proofs can either appear in the main paper or the supplemental material, but if they appear in the supplemental material, the authors are encouraged to provide a short proof sketch to provide intuition. 
        \item Inversely, any informal proof provided in the core of the paper should be complemented by formal proofs provided in appendix or supplemental material.
        \item Theorems and Lemmas that the proof relies upon should be properly referenced. 
    \end{itemize}

    \item {\bf Experimental result reproducibility}
    \item[] Question: Does the paper fully disclose all the information needed to reproduce the main experimental results of the paper to the extent that it affects the main claims and/or conclusions of the paper (regardless of whether the code and data are provided or not)?
    \item[] Answer: \answerTODO{} % Replace by \answerYes{}, \answerNo{}, or \answerNA{}.
    \item[] Justification: \justificationTODO{}
    \item[] Guidelines:
    \begin{itemize}
        \item The answer \answerNA{} means that the paper does not include experiments.
        \item If the paper includes experiments, a \answerNo{} answer to this question will not be perceived well by the reviewers: Making the paper reproducible is important, regardless of whether the code and data are provided or not.
        \item If the contribution is a dataset and\slash or model, the authors should describe the steps taken to make their results reproducible or verifiable. 
        \item Depending on the contribution, reproducibility can be accomplished in various ways. For example, if the contribution is a novel architecture, describing the architecture fully might suffice, or if the contribution is a specific model and empirical evaluation, it may be necessary to either make it possible for others to replicate the model with the same dataset, or provide access to the model. In general. releasing code and data is often one good way to accomplish this, but reproducibility can also be provided via detailed instructions for how to replicate the results, access to a hosted model (e.g., in the case of a large language model), releasing of a model checkpoint, or other means that are appropriate to the research performed.
        \item While NeurIPS does not require releasing code, the conference does require all submissions to provide some reasonable avenue for reproducibility, which may depend on the nature of the contribution. For example
        \begin{enumerate}
            \item If the contribution is primarily a new algorithm, the paper should make it clear how to reproduce that algorithm.
            \item If the contribution is primarily a new model architecture, the paper should describe the architecture clearly and fully.
            \item If the contribution is a new model (e.g., a large language model), then there should either be a way to access this model for reproducing the results or a way to reproduce the model (e.g., with an open-source dataset or instructions for how to construct the dataset).
            \item We recognize that reproducibility may be tricky in some cases, in which case authors are welcome to describe the particular way they provide for reproducibility. In the case of closed-source models, it may be that access to the model is limited in some way (e.g., to registered users), but it should be possible for other researchers to have some path to reproducing or verifying the results.
        \end{enumerate}
    \end{itemize}

\item {\bf Open access to data and code}
    \item[] Question: Does the paper provide open access to the data and code, with sufficient instructions to faithfully reproduce the main experimental results, as described in supplemental material?
    \item[] Answer: \answerTODO{} % Replace by \answerYes{}, \answerNo{}, or \answerNA{}.
    \item[] Justification: \justificationTODO{}
    \item[] Guidelines:
    \begin{itemize}
        \item The answer \answerNA{} means that paper does not include experiments requiring code.
        \item Please see the NeurIPS code and data submission guidelines (\url{https://neurips.cc/public/guides/CodeSubmissionPolicy}) for more details.
        \item While we encourage the release of code and data, we understand that this might not be possible, so \answerNo{} is an acceptable answer. Papers cannot be rejected simply for not including code, unless this is central to the contribution (e.g., for a new open-source benchmark).
        \item The instructions should contain the exact command and environment needed to run to reproduce the results. See the NeurIPS code and data submission guidelines (\url{https://neurips.cc/public/guides/CodeSubmissionPolicy}) for more details.
        \item The authors should provide instructions on data access and preparation, including how to access the raw data, preprocessed data, intermediate data, and generated data, etc.
        \item The authors should provide scripts to reproduce all experimental results for the new proposed method and baselines. If only a subset of experiments are reproducible, they should state which ones are omitted from the script and why.
        \item At submission time, to preserve anonymity, the authors should release anonymized versions (if applicable).
        \item Providing as much information as possible in supplemental material (appended to the paper) is recommended, but including URLs to data and code is permitted.
    \end{itemize}

\item {\bf Experimental setting/details}
    \item[] Question: Does the paper specify all the training and test details (e.g., data splits, hyperparameters, how they were chosen, type of optimizer) necessary to understand the results?
    \item[] Answer: \answerTODO{} % Replace by \answerYes{}, \answerNo{}, or \answerNA{}.
    \item[] Justification: \justificationTODO{}
    \item[] Guidelines:
    \begin{itemize}
        \item The answer \answerNA{} means that the paper does not include experiments.
        \item The experimental setting should be presented in the core of the paper to a level of detail that is necessary to appreciate the results and make sense of them.
        \item The full details can be provided either with the code, in appendix, or as supplemental material.
    \end{itemize}

\item {\bf Experiment statistical significance}
    \item[] Question: Does the paper report error bars suitably and correctly defined or other appropriate information about the statistical significance of the experiments?
    \item[] Answer: \answerTODO{} % Replace by \answerYes{}, \answerNo{}, or \answerNA{}.
    \item[] Justification: \justificationTODO{}
    \item[] Guidelines:
    \begin{itemize}
        \item The answer \answerNA{} means that the paper does not include experiments.
        \item The authors should answer \answerYes{} if the results are accompanied by error bars, confidence intervals, or statistical significance tests, at least for the experiments that support the main claims of the paper.
        \item The factors of variability that the error bars are capturing should be clearly stated (for example, train/test split, initialization, random drawing of some parameter, or overall run with given experimental conditions).
        \item The method for calculating the error bars should be explained (closed form formula, call to a library function, bootstrap, etc.)
        \item The assumptions made should be given (e.g., Normally distributed errors).
        \item It should be clear whether the error bar is the standard deviation or the standard error of the mean.
        \item It is OK to report 1-sigma error bars, but one should state it. The authors should preferably report a 2-sigma error bar than state that they have a 96\% CI, if the hypothesis of Normality of errors is not verified.
        \item For asymmetric distributions, the authors should be careful not to show in tables or figures symmetric error bars that would yield results that are out of range (e.g., negative error rates).
        \item If error bars are reported in tables or plots, the authors should explain in the text how they were calculated and reference the corresponding figures or tables in the text.
    \end{itemize}

\item {\bf Experiments compute resources}
    \item[] Question: For each experiment, does the paper provide sufficient information on the computer resources (type of compute workers, memory, time of execution) needed to reproduce the experiments?
    \item[] Answer: \answerTODO{} % Replace by \answerYes{}, \answerNo{}, or \answerNA{}.
    \item[] Justification: \justificationTODO{}
    \item[] Guidelines:
    \begin{itemize}
        \item The answer \answerNA{} means that the paper does not include experiments.
        \item The paper should indicate the type of compute workers CPU or GPU, internal cluster, or cloud provider, including relevant memory and storage.
        \item The paper should provide the amount of compute required for each of the individual experimental runs as well as estimate the total compute. 
        \item The paper should disclose whether the full research project required more compute than the experiments reported in the paper (e.g., preliminary or failed experiments that didn't make it into the paper). 
    \end{itemize}
    
\item {\bf Code of ethics}
    \item[] Question: Does the research conducted in the paper conform, in every respect, with the NeurIPS Code of Ethics \url{https://neurips.cc/public/EthicsGuidelines}?
    \item[] Answer: \answerTODO{} % Replace by \answerYes{}, \answerNo{}, or \answerNA{}.
    \item[] Justification: \justificationTODO{}
    \item[] Guidelines:
    \begin{itemize}
        \item The answer \answerNA{} means that the authors have not reviewed the NeurIPS Code of Ethics.
        \item If the authors answer \answerNo, they should explain the special circumstances that require a deviation from the Code of Ethics.
        \item The authors should make sure to preserve anonymity (e.g., if there is a special consideration due to laws or regulations in their jurisdiction).
    \end{itemize}

\item {\bf Broader impacts}
    \item[] Question: Does the paper discuss both potential positive societal impacts and negative societal impacts of the work performed?
    \item[] Answer: \answerTODO{} % Replace by \answerYes{}, \answerNo{}, or \answerNA{}.
    \item[] Justification: \justificationTODO{}
    \item[] Guidelines:
    \begin{itemize}
        \item The answer \answerNA{} means that there is no societal impact of the work performed.
        \item If the authors answer \answerNA{} or \answerNo, they should explain why their work has no societal impact or why the paper does not address societal impact.
        \item Examples of negative societal impacts include potential malicious or unintended uses (e.g., disinformation, generating fake profiles, surveillance), fairness considerations (e.g., deployment of technologies that could make decisions that unfairly impact specific groups), privacy considerations, and security considerations.
        \item The conference expects that many papers will be foundational research and not tied to particular applications, let alone deployments. However, if there is a direct path to any negative applications, the authors should point it out. For example, it is legitimate to point out that an improvement in the quality of generative models could be used to generate Deepfakes for disinformation. On the other hand, it is not needed to point out that a generic algorithm for optimizing neural networks could enable people to train models that generate Deepfakes faster.
        \item The authors should consider possible harms that could arise when the technology is being used as intended and functioning correctly, harms that could arise when the technology is being used as intended but gives incorrect results, and harms following from (intentional or unintentional) misuse of the technology.
        \item If there are negative societal impacts, the authors could also discuss possible mitigation strategies (e.g., gated release of models, providing defenses in addition to attacks, mechanisms for monitoring misuse, mechanisms to monitor how a system learns from feedback over time, improving the efficiency and accessibility of ML).
    \end{itemize}
    
\item {\bf Safeguards}
    \item[] Question: Does the paper describe safeguards that have been put in place for responsible release of data or models that have a high risk for misuse (e.g., pre-trained language models, image generators, or scraped datasets)?
    \item[] Answer: \answerTODO{} % Replace by \answerYes{}, \answerNo{}, or \answerNA{}.
    \item[] Justification: \justificationTODO{}
    \item[] Guidelines:
    \begin{itemize}
        \item The answer \answerNA{} means that the paper poses no such risks.
        \item Released models that have a high risk for misuse or dual-use should be released with necessary safeguards to allow for controlled use of the model, for example by requiring that users adhere to usage guidelines or restrictions to access the model or implementing safety filters. 
        \item Datasets that have been scraped from the Internet could pose safety risks. The authors should describe how they avoided releasing unsafe images.
        \item We recognize that providing effective safeguards is challenging, and many papers do not require this, but we encourage authors to take this into account and make a best faith effort.
    \end{itemize}

\item {\bf Licenses for existing assets}
    \item[] Question: Are the creators or original owners of assets (e.g., code, data, models), used in the paper, properly credited and are the license and terms of use explicitly mentioned and properly respected?
    \item[] Answer: \answerTODO{} % Replace by \answerYes{}, \answerNo{}, or \answerNA{}.
    \item[] Justification: \justificationTODO{}
    \item[] Guidelines:
    \begin{itemize}
        \item The answer \answerNA{} means that the paper does not use existing assets.
        \item The authors should cite the original paper that produced the code package or dataset.
        \item The authors should state which version of the asset is used and, if possible, include a URL.
        \item The name of the license (e.g., CC-BY 4.0) should be included for each asset.
        \item For scraped data from a particular source (e.g., website), the copyright and terms of service of that source should be provided.
        \item If assets are released, the license, copyright information, and terms of use in the package should be provided. For popular datasets, \url{paperswithcode.com/datasets} has curated licenses for some datasets. Their licensing guide can help determine the license of a dataset.
        \item For existing datasets that are re-packaged, both the original license and the license of the derived asset (if it has changed) should be provided.
        \item If this information is not available online, the authors are encouraged to reach out to the asset's creators.
    \end{itemize}

\item {\bf New assets}
    \item[] Question: Are new assets introduced in the paper well documented and is the documentation provided alongside the assets?
    \item[] Answer: \answerTODO{} % Replace by \answerYes{}, \answerNo{}, or \answerNA{}.
    \item[] Justification: \justificationTODO{}
    \item[] Guidelines:
    \begin{itemize}
        \item The answer \answerNA{} means that the paper does not release new assets.
        \item Researchers should communicate the details of the dataset\slash code\slash model as part of their submissions via structured templates. This includes details about training, license, limitations, etc. 
        \item The paper should discuss whether and how consent was obtained from people whose asset is used.
        \item At submission time, remember to anonymize your assets (if applicable). You can either create an anonymized URL or include an anonymized zip file.
    \end{itemize}

\item {\bf Crowdsourcing and research with human subjects}
    \item[] Question: For crowdsourcing experiments and research with human subjects, does the paper include the full text of instructions given to participants and screenshots, if applicable, as well as details about compensation (if any)? 
    \item[] Answer: \answerTODO{} % Replace by \answerYes{}, \answerNo{}, or \answerNA{}.
    \item[] Justification: \justificationTODO{}
    \item[] Guidelines:
    \begin{itemize}
        \item The answer \answerNA{} means that the paper does not involve crowdsourcing nor research with human subjects.
        \item Including this information in the supplemental material is fine, but if the main contribution of the paper involves human subjects, then as much detail as possible should be included in the main paper. 
        \item According to the NeurIPS Code of Ethics, workers involved in data collection, curation, or other labor should be paid at least the minimum wage in the country of the data collector. 
    \end{itemize}

\item {\bf Institutional review board (IRB) approvals or equivalent for research with human subjects}
    \item[] Question: Does the paper describe potential risks incurred by study participants, whether such risks were disclosed to the subjects, and whether Institutional Review Board (IRB) approvals (or an equivalent approval/review based on the requirements of your country or institution) were obtained?
    \item[] Answer: \answerTODO{} % Replace by \answerYes{}, \answerNo{}, or \answerNA{}.
    \item[] Justification: \justificationTODO{}
    \item[] Guidelines:
    \begin{itemize}
        \item The answer \answerNA{} means that the paper does not involve crowdsourcing nor research with human subjects.
        \item Depending on the country in which research is conducted, IRB approval (or equivalent) may be required for any human subjects research. If you obtained IRB approval, you should clearly state this in the paper. 
        \item We recognize that the procedures for this may vary significantly between institutions and locations, and we expect authors to adhere to the NeurIPS Code of Ethics and the guidelines for their institution. 
        \item For initial submissions, do not include any information that would break anonymity (if applicable), such as the institution conducting the review.
    \end{itemize}

\item {\bf Declaration of LLM usage}
    \item[] Question: Does the paper describe the usage of LLMs if it is an important, original, or non-standard component of the core methods in this research? Note that if the LLM is used only for writing, editing, or formatting purposes and does \emph{not} impact the core methodology, scientific rigor, or originality of the research, declaration is not required.
    %this research? 
    \item[] Answer: \answerTODO{} % Replace by \answerYes{}, \answerNo{}, or \answerNA{}.
    \item[] Justification: \justificationTODO{}
    \item[] Guidelines:
    \begin{itemize}
        \item The answer \answerNA{} means that the core method development in this research does not involve LLMs as any important, original, or non-standard components.
        \item Please refer to our LLM policy in the NeurIPS handbook for what should or should not be described.
    \end{itemize}

\end{enumerate}
\fi

\end{document}